\documentclass[10pt]{article}

\usepackage[T1]{fontenc}
\usepackage{lmodern}
\usepackage[british]{babel}

\usepackage{mathtools}
\usepackage{amssymb}
\usepackage{amsthm}

\usepackage{graphicx}
\usepackage{float}
\usepackage{caption}
\usepackage{parskip}

\usepackage{geometry}
\usepackage[unicode]{hyperref}

\newtheorem{rem}{Remark}

\begin{document}

\title{Optimal Quoting under Adverse Selection and Price Reading\footnote{A shortened version of this paper is to be published in Risk (Cutting Edge) in 2026.}}
\author{Alexander \textsc{Barzykin}\footnote{HSBC, 8 Canada Square, Canary Wharf, London E14 5HQ, United Kingdom, \texttt{alexander.barzykin@hsbc.com}.} \and Philippe \textsc{Bergault}\footnote{Université Paris Dauphine-PSL, CEREMADE, 75775 Paris Cedex 16, France, \texttt{bergault@ceremade.dauphine.fr}.} \and Olivier \textsc{Guéant}\footnote{Université Paris Cité, Laboratoire de Probabilités, Statistique et Modélisation, 75205 Paris Cedex 13, France, \texttt{olivier.gueant@u-pariscite.fr}.} \and Malo \textsc{Lemmel}\footnote{EPFL student and CNRS intern, \texttt{malo.lemmel@epfl.ch}.}}
\date{}
\maketitle

\begin{abstract}
\noindent
Over the past decade, many dealers have implemented algorithmic models to automatically respond to RFQs and manage flows originating from electronic platforms. In parallel, building on the foundational work of Ho and Stoll, and later Avellaneda and Stoikov, the academic literature on market making has expanded to address trade size distributions, client tiering, complex price dynamics, alpha signals and the internalisation versus externalisation dilemma in markets with dealer-to-client and interdealer-broker segments. In this paper, we tackle two critical dimensions: adverse selection, arising from the presence of informed traders, and price reading, whereby the market maker's own quotes reveal the direction of their inventory. These risks are well known to practitioners, who routinely face informed flows and algorithms capable of extracting signals from quoting behaviour. Yet they have received limited attention in the quantitative finance literature, beyond stylised toy models with limited actionability. Extending the existing literature, we propose a tractable framework that enables market makers to adjust their quotes with greater awareness of informational risk.
\end{abstract}

\vspace{7mm}


\vspace{5mm}

\section{Introduction}

In modern dealer markets, liquidity provision is increasingly handled by algorithmic systems that autonomously respond to clients' RFQs and manage trade flows across multiple platforms. The academic literature has accompanied this technological shift with increasingly refined models. Building on the seminal contributions of Ho and Stoll~\cite{ho1981optimal} and Avellaneda and Stoikov~\cite{avellaneda2008high}, researchers have developed several models based on optimal control to deliver bid and ask price ladders to various types of clients, while managing inventory risk through quote skewing (internalisation) and, when possible, hedging on trading venues (externalisation); see~\cite{barzykin2023algorithmic, butz2019, gueant2016financial, gueant2013dealing}.

Some models feature complex price dynamics \cite{rosenbaum2022}, others incorporate alpha signals; see \cite{cartea2015algorithmic, cartea2014buy}. 
Several focus on multi-asset market making under correlation or cointegration assumptions; see \cite{barzykin2023dealing, barzykin2024market}. On the numerical side, substantial progress has been made, and reference methods now enable the approximation of optimal quotes in many contexts without suffering from the curse of dimensionality; see \cite{bergault2021closed}.

Yet, despite their sophistication, most existing market making models overlook two key challenges that are central to real-world practice: adverse selection and information leakage through price reading. Adverse selection arises when the market maker trades against participants whose order flow reflects superior information about future price dynamics. In other words, the market maker faces the risk of being hit at the bid (respectively ask) just before the market goes down (respectively up) -- the so-called winner's curse. Information leakage, as captured here by price reading, occurs when the market maker's own quotes indirectly reveal information about their inventory or at least its sign -- allowing certain traders, often informally referred to as skew sniffers (see \cite{goyder2024}), to anticipate future market flows and subsequently build potentially profitable strategies.

While these risks are well known to practitioners (see \cite{baldauf2024}), they are rarely addressed in the quantitative finance literature. When they are, it is typically through highly stylised economic models, offering little guidance for implementation; see for instance~\cite{glosten1985bid}.\footnote{Among the exceptions, \cite{cartea2025simple} is an interesting paper helpful in addressing the question of toxic flow, intimately related to our question of adverse selection.} In particular, the interplay between informational risk and real-time quoting decisions remains insufficiently studied.

The goal of this paper is to help fill that gap. We introduce a tractable optimal market making model \textit{à la} Cartea-Jaimungal, in infinite horizon, that accounts for both adverse selection and price reading effects. To avoid introducing a lot of assumptions, we rely on Taylor expansions and focus on the first-order adjustments in optimal quotes induced by each effect. The resulting equations provide practical insight into how market makers can adapt their behaviour to flow toxicity and mitigate the risk of information leakage, while remaining competitive.

The remainder of the paper is organised as follows. Section~\ref{sec:genpb} introduces the model. In Section~\ref{sec:hjb_fo}, we derive the first-order effects of adverse selection and price reading on both the value function and the optimal quotes. Section~\ref{sec:begv} discusses informational effects in the context of a quadratic approximation of the Hamiltonian functions. Section~\ref{sec:numerics} illustrates our findings in the case of simple functional forms. Section \ref{sec:conclusion} concludes.

\section{Model Setup}
\label{sec:genpb}

Throughout this paper, we work on a filtered probability space \(
\left( \Omega, \mathcal{F}, \mathbb{P}; \mathbb{F} = \left( \mathcal{F}_{t} \right)_{t \geq 0} \right)
\) satisfying the usual conditions. We assume that this space is rich enough to support all the stochastic processes introduced in the model.

\subsection{Modelling Framework and Notation}

We consider a market maker responsible for quoting bid and ask prices on a single asset. The market maker continuously sets quotes and can stream different price ladders to different categories of clients. Specifically, clients are divided into \( N \) tiers, indexed by the set \( \mathcal{N} := \{1, \ldots, N\} \).

Transaction sizes are assumed to be discrete, taking positive values \( \Delta^1 < \ldots < \Delta^K \) for some \( K \geq 1 \), with index set \( \mathcal{K} := \{1, \ldots, K\} \). We assume \( \Delta^k \in \Delta^1 \mathbb{N}^* \) for all \(k \in \mathcal K\), i.e., each size is an integer multiple of the smallest one.

For each tier \( n \in \mathcal{N} \) and size index \( k \in \mathcal K \), the bid and ask prices quoted at time \( t \) are modelled by predictable stochastic processes \( \big(S^{n,k,b}_t\big)_{t \geq 0} \) and \( \big(S^{n,k,a}_t\big)_{t \geq 0} \), respectively. These quotes determine the distribution of trade arrivals, which will be formalised later.

We introduce a reference price process \( \big(S_t\big)_{t \geq 0} \), representing a composite price computed by the market maker based on public and private information. It evolves under both exogenous volatility and endogenous reactions to trading activity and quote asymmetry. Its dynamics is given by:
\begin{align}
\label{eq:refPrice}
dS_t =\ & \sigma\, dB_t 
- \sum_{\substack{n \in \mathcal{N} \\ k \in \mathcal{K}}} \left(
\tilde{\zeta}^{n,k}\left(S_{t-} - S^{n,k,b}_t\right)\, dN^{n,k,b}_t
- \tilde{\zeta}^{n,k}\left(S^{n,k,a}_t - S_{t-}\right)\, dN^{n,k,a}_t
\right) \\
& + \sum_{n \in \mathcal{N}} \tilde{J}^n\left( \sum_{k \in \mathcal{K}} w^{n,k} \left( S^{n,k,a}_t - 2 S_{t-} + S^{n,k,b}_t \right) \right)\, dt, \qquad S_0 \text{ given,} \nonumber
\end{align}
where \( \sigma > 0 \) is a volatility parameter and \( \big(B_t\big)_{t \geq 0} \) is a standard Brownian motion adapted to the filtration \( \left(\mathcal{F}_t\right)_{t \geq 0} \). The point processes \( \big(N^{n,k,b}_t\big)_{t \geq 0} \) and \( \big(N^{n,k,a}_t\big)_{t \geq 0} \) represent the number of trades of size \( \Delta^k \) at the bid and ask with clients from tier \( n \), respectively.

The model includes two key features that govern the endogenous reaction of the reference price to trade flow and quoting behaviour:

\begin{itemize}
	\item \textbf{Adverse selection.} For each \( (n,k) \in \mathcal{N} \times \mathcal{K} \), the function \( \tilde{\zeta}^{n,k} : \mathbb{R} \to \mathbb{R}_+ \) is nondecreasing, and satisfies
\(
\lim_{\delta \to -\infty} \tilde{\zeta}^{n,k}(\delta) = 0.
\)
This function captures the impact of informed trading on prices. It is tier-specific (and possibly size-specific), allowing the market maker to account for varying levels of expected informativeness across clients (and transaction sizes). In particular, when \(\tilde{\zeta}^{n,k}\) is a rapidly increasing function, executions at prices seemingly unfavourable to the client, from the market maker's perspective, can have a strong impact on the reference price. The dependence on the quote offset should be interpreted conditionally on execution. A transaction at a conservative quote is less likely, but, conditional on such a transaction occurring, it may reveal a private signal, a stale quote and/or a speed advantage. When this conditional impact effect is not empirically supported, the constant specification \(\tilde\zeta^{n,k}(\delta)=\tilde\zeta^{n,k}\) is a natural choice; see \cite{gueant2013dealing}.

\item \textbf{Price reading.} For each \( n \in \mathcal{N} \), the function \( \tilde{J}^n : \mathbb{R} \to \mathbb{R} \) is nondecreasing and satisfies \( \tilde{J}^n(0) = 0 \). It models how skew sniffers from tier \(n\) drive the reference price in reaction to asymmetry in the quotes streamed to them. This asymmetry, or skew, is computed as a weighted sum over the quote skew for each trade size~\( \Delta^k \), where the inner term \( S^{n,k,a}_t - 2S_{t-} + S^{n,k,b}_t \) captures the relative placement of bid and ask quotes with respect to the reference price. The (nonnegative) weights \( w^{n,\cdot} \) control the contribution of each size in the price ladder to the total skew associated with the quotes streamed to clients of tier \( n \). This mechanism captures the phenomenon commonly referred to as price reading or skew sniffing, in which certain market participants infer the direction -- or evolution -- of the market maker's inventory from the quote ladder structure and adjust their trading behaviour accordingly, generating a feedback loop that moves the reference price.
\end{itemize}

\begin{rem}
Throughout, \(\tilde{J}^n\) and \(\tilde{\zeta}^{n,k}\) are local descriptions of the
informational environment around the quotes prevailing in the absence of
informational risk: our analysis involves them only through their values and
first derivatives at those baseline quotes, and the qualitative properties
assumed above do not, on their own, guarantee that the optimisation problems of
Section~\ref{sec:hjb_fo} are well posed for arbitrary quotes. This is immaterial
for the perturbative analysis carried out here.
\end{rem}

\begin{rem}
In the case of price reading,
supply-demand imbalance universally appears in short-term price prediction
models (see \cite{lucchese2024} and \cite{sirignano2019}), so any skew can be
interpreted as a signal; the price reaction should therefore be based on
competitiveness metrics rather than on raw quote skews. The Appendix presents
such a bounded competitiveness-based variant, in which only quotes close enough
to the market contribute significantly to price reading. In the context of
multi-dealer platforms, only the competitive side may be of importance.
\end{rem}

The market maker's inventory process \( \left(q_t\right)_{t \geq 0} \) evolves according to
\begin{equation}
\label{eq:inventory}
dq_t = \sum_{\substack{n \in \mathcal{N} \\ k \in \mathcal{K}}} \Delta^k \left( dN^{n,k,b}_t - dN^{n,k,a}_t \right), \quad q_0 \text{ given}.
\end{equation}

The processes \( \left( N^{n,k,b}_t \right)_{t \geq 0} \) and \( \left( N^{n,k,a}_t \right)_{t \geq 0} \) have respective intensities \( \left( \lambda^{n,k,b}_t \right)_{t \geq 0} \) and \( \left( \lambda^{n,k,a}_t \right)_{t \geq 0} \), defined by
\begin{equation}
\label{eq:intensity}
\lambda^{n,k,b}_t = \Lambda^{n,k,b} \left( S_{t-} - S^{n,k,b}_t\right) \quad \text{and} \quad
\lambda^{n,k,a}_t = \Lambda^{n,k,a} \left( S^{n,k,a}_t - S_{t-} \right).
\end{equation}

Here, \( \Lambda^{n,k,b} \) and \( \Lambda^{n,k,a} \) are smooth, positive and decreasing functions, satisfying standard assumptions in market making models \textit{à la} Avellaneda-Stoikov or Cartea-Jaimungal (see~\cite{gueant2016financial}); in particular, 
\[
c^{n,k,b/a}(\delta) := 2 - \frac{\Lambda^{n,k,b/a}(\delta) \Lambda^{n,k,b/a\prime\prime}(\delta)}{\Lambda^{n,k,b/a\prime}(\delta)^2}
\] 
verifies \(\inf_\delta c^{n,k,b/a}(\delta) >0\).

From a calibration perspective, the functions \(\Lambda^{n,k,b/a}\) are intensity curves estimated from RFQ and execution data conditional on tier, size and side. The adverse-selection functions \(\tilde{\zeta}^{n,k}\) summarise post-fill markouts of the reference price, conditional on execution at a given offset. The price-reading loadings \(w^{n,k}\) and response functions \(\tilde{J}^n\) summarise how much information clients from a given tier extract from displayed skew, and may be proxied by tier-level analysis such as skew sensitivity, subsequent trading direction, or observed deterioration in post-quote markouts. These quantities are treated as inputs in the theoretical analysis; estimating them robustly is an implementation step rather than part of the current model.

\begin{rem}
Although the reference price -- whose dynamics is given by Eq.~\eqref{eq:refPrice} -- is primarily driven by its exogenous Brownian component, it also depends on the market maker's clients' flows and own quoting behaviour. As such, it is a price specific to the market maker under consideration. In light of this, it may seem questionable at first glance to define the skew terms in Eq.~\eqref{eq:refPrice} relative to this very reference price, and to let the intensities in Eq.~\eqref{eq:intensity} -- which model the agents' demand and supply behaviour -- also depend on the market maker's own reference price. This modelling choice is, in fact, a conservative one: we assume that the market maker never benefits, in an exclusive way, from any informational advantage, even though they observe clients' flows. While restrictive, this assumption appears natural in the context of a paper focused on the informational risk faced by market makers. It is recognised that adverse selection can originate from the market maker not knowing the true reference price at the right time due to natural market fragmentation and the ensuing latency.
\end{rem}

The market maker's cash process \( \left( X_t \right)_{t \geq 0} \) evolves according to:
\begin{equation}
\label{eq:cash}
\begin{aligned}
dX_t 
&= \sum_{\substack{n \in \mathcal{N} \\ k \in \mathcal{K}}} \Delta^k \left(- S^{n,k,b}_t\, dN^{n,k,b}_t + S^{n,k,a}_t\, dN^{n,k,a}_t  \right) \\
&= \sum_{\substack{n \in \mathcal{N} \\ k \in \mathcal{K}}} \Delta^k \left(- \left( S_{t-} - \delta^{n,k,b}_t \right)\, dN^{n,k,b}_t + \left( S_{t-} + \delta^{n,k,a}_t \right)\, dN^{n,k,a}_t \right) \\
&= -S_{t-}\, dq_t + \sum_{\substack{n \in \mathcal{N} \\ k \in \mathcal{K}}} \Delta^k \left( \delta^{n,k,b}_t\, dN^{n,k,b}_t + \delta^{n,k,a}_t\, dN^{n,k,a}_t  \right),
\end{aligned}
\end{equation}
where \( \delta^{n,k,b}_t := S_{t-} - S^{n,k,b}_t \) and \( \delta^{n,k,a}_t := S^{n,k,a}_t - S_{t-} \) are the bid and ask quote offsets -- that we often refer to, by abuse of language, as the bid and ask quotes. They represent the markdown and markup relative to the reference price just before time \(t\), i.e. \( S_{t-} \).

\subsection{The Optimisation Problem}

In the above discussion, we introduced the three processes central to most market making models: the reference price process \( \left(S_t\right)_{t \geq 0} \), the inventory process \( \left(q_t\right)_{t \geq 0} \) and the cash process \( \left(X_t\right)_{t \geq 0} \).

Assuming that the market maker continuously marks their inventory to the reference price, the dynamics of the profit and loss process \( \left(\mathrm{PnL}_t\right)_{t \geq 0} \) of the market maker can be computed from Eqs. \eqref{eq:refPrice}, \eqref{eq:inventory} and \eqref{eq:cash}:
\begin{align*}
d\mathrm{PnL}_t &= dX_t + S_{t-}\, dq_t + q_{t-}\, dS_t + d[ q, S]_t \\
&= \sum_{\substack{n \in \mathcal{N} \\ k \in \mathcal{K}}} \left(
\big( \Delta^k \delta^{n,k,b}_t - (q_{t-} + \Delta^k)\, \tilde{\zeta}^{n,k}(\delta^{n,k,b}_t) \big) dN^{n,k,b}_t + \big( \Delta^k \delta^{n,k,a}_t + (q_{t-} - \Delta^k)\, \tilde{\zeta}^{n,k}(\delta^{n,k,a}_t) \big) dN^{n,k,a}_t \right) \\
&\quad + q_{t-}\, \sigma\, dB_t + q_{t-} \sum_{n \in \mathcal{N}} \tilde{J}^n \left( \sum_{k \in \mathcal{K}} w^{n,k} \left( \delta^{n,k,a}_t - \delta^{n,k,b}_t \right) \right) dt.
\end{align*}

A classical optimisation approach is then to consider a finite-horizon objective, where the market maker maximises the expected PnL at a given time \( T \), while penalising inventory risk over the interval \([0,T]\). This leads to the following risk-adjusted expected PnL objective, common in the works of Cartea, Jaimungal and coauthors (see e.g.~\cite{cartea2015algorithmic}):
\[
\mathbb{E}\left[\mathrm{PnL}_T - \frac{\gamma}{2}\int_{0}^{T} \sigma^2 q_t^2\, dt\right],
\]
maximised over the family of predictable and bounded-from-below control processes \(\left( \left(\delta_t^{n,k,b},\delta_t^{n,k,a} \right)_{t\geq 0} \right)_{(n,k) \in \mathcal{N} \times \mathcal{K}}\).

Here and below, \(\gamma>0\) denotes the market maker's risk aversion coefficient.
See also~\cite{gueant2016financial} for a broader perspective on market making objectives.

In this paper, we instead adopt a stationary infinite-horizon perspective,\footnote{In market making models, the focus naturally lies on the optimal behaviour far from the terminal time (which is often arbitrary in OTC markets). In the analysis that follows, we use Taylor expansions to capture the first-order impact of informational risk, working directly in an infinite-horizon framework. Performing these expansions in a finite-horizon setting and then taking the limit as \(T \to +\infty\) would lead to results with no financial relevance (this is the classical iterated limit problem).} and consider the discounted objective:
\begin{equation*}
\mathbb{E}\left[ \int_0^{+\infty} e^{-\rho t} \left( d\mathrm{PnL}_t - \frac{\gamma}{2} \sigma^2 q_t^2\, dt \right) \right],
\end{equation*}
where \( \rho > 0 \) is a discount rate.

Using the form of intensities given by Eq. \eqref{eq:intensity}, this objective is equivalent to
\begin{equation}
\label{eq:LTproblem}
\mathbb{E}\left[ \int_0^{+\infty} e^{-\rho t} \left(
\sum_{n \in \mathcal{N}} \mathfrak{b}^n\left(q_{t-}, \delta^{n,\cdot,b}_t, \delta^{n,\cdot,a}_t \right) 
- \frac{\gamma}{2} \sigma^2 q_t^2 \right) dt \right],
\end{equation}
where for each \( n \in \mathcal{N} \), the function \( \mathfrak{b}^n \) is defined as:
\begin{align}
\mathfrak{b}^n\Big( q, \delta^{n,\cdot,b} ,\delta^{n,\cdot,a} \Big) 
&= q\, \tilde{J}^n\left( \sum_{k \in \mathcal{K}} w^{n,k} \left( \delta^{n,k,a} - \delta^{n,k,b} \right) \right) \nonumber \\
&\quad + \sum_{k \in \mathcal{K}} \Bigg( \Delta^k\, \Lambda^{n,k,b}\left( \delta^{n,k,b} \right) 
\left( \delta^{n,k,b} - \frac{q + \Delta^k}{\Delta^k} \tilde{\zeta}^{n,k}\left( \delta^{n,k,b} \right) \right) \nonumber \\
&\qquad \qquad + \Delta^k\, \Lambda^{n,k,a}\left( \delta^{n,k,a} \right) 
\left( \delta^{n,k,a} + \frac{q - \Delta^k}{\Delta^k} \tilde{\zeta}^{n,k}\left( \delta^{n,k,a} \right) \right) \Bigg). \nonumber
\end{align}

\section{Informational Risk: A First-Order Perspective}\label{sec:hjb_fo}

The aim of this section is to understand the influence of informational risk on the optimal behaviour of a market maker. To this end, we analyse the optimal control problem associated with the objective function in Eq.~\eqref{eq:LTproblem}. Rather than solving the problem fully, we aim to characterise the first-order effects of informational risk. The results should therefore be interpreted as local comparative statics: they are most reliable when the informational risk is small and the adjusted quotes remain close to the baseline policy. Interactions between adverse selection and price reading only appear beyond first order and are not captured by the formulas in this section. We begin by presenting the Hamilton-Jacobi-Bellman (HJB) equation associated with the general model, then recall classical results in the absence of informational risk, and finally compute first-order adjustments.

\subsection{The Hamilton-Jacobi-Bellman Equation}

The HJB equation corresponding to the infinite horizon problem~\eqref{eq:LTproblem} is a nonlinear system for the value function
\( \vartheta : \Delta^1 \mathbb Z \to \mathbb{R} \), given by:
\begin{align}
\label{eq:thetaModel_main}
0 &= -\rho\, \vartheta(q) - \frac{\gamma}{2} \sigma^2 q^2
+ \sum_{n \in \mathcal{N}} \tilde{\mathcal{H}}^n\Bigl[\tilde{J}^n , \tilde{\zeta}^{n,\cdot} \Bigr]\left(q, D^\cdot_+ \vartheta(q), D^\cdot_- \vartheta(q) \right),
\end{align}
where the finite-difference operators are defined, for all \(k \in \mathcal{K}\), by
\[
D^k_+ \vartheta(q) := \frac{\vartheta(q) - \vartheta(q + \Delta^k)}{\Delta^k}  \quad \text{and} \quad 
D^k_- \vartheta(q) := \frac{\vartheta(q) - \vartheta(q - \Delta^k)}{\Delta^k},
\]
and for each \( n \in \mathcal{N} \), the Hamiltonian \( \tilde{\mathcal{H}}^n \) is given by
\begin{align*}
&\tilde{\mathcal{H}}^n\left[ \tilde{J}^n , \tilde{\zeta}^{n,\cdot} \right]
\left( q, p^{\cdot,b}, p^{\cdot,a} \right) := \  \sup_{\delta^{n,\cdot,b}, \delta^{n,\cdot,a} } \Bigg\{
\mathfrak{b}^n\left( q, \delta^{n,\cdot,b}, \delta^{n,\cdot,a} \right)\\ 
& \qquad\qquad\qquad\qquad\qquad - \sum_{k \in \mathcal{K}} \Delta^k \left(
\Lambda^{n,k,b}(\delta^{n,k,b})\, p^{k,b} +
\Lambda^{n,k,a}(\delta^{n,k,a})\, p^{k,a} \right)
\Bigg\}.
\end{align*}

Whenever the above suprema are attained, we denote by
\(\tilde{\mathfrak{d}}^{n,\cdot,b/a*}(q,p^{\cdot,b},p^{\cdot,a})\) the
corresponding maximisers, and, under mild assumptions, the optimal quotes for tier \(n\) in feedback
form, denoted by \(\mathfrak{d}^{n,\cdot,b/a*}(q)\), are then given by
\(\tilde{\mathfrak{d}}^{n,\cdot,b/a*}\left(q, D^\cdot_+ \vartheta(q),
D^\cdot_- \vartheta(q) \right)\). Consistently with the local reading of
\(\tilde{J}^n\) and \(\tilde{\zeta}^{n,\cdot}\) discussed in
Section~\ref{sec:genpb}, we do not investigate the conditions under which
this holds globally: the analysis that follows is perturbative around the
baseline of Section~\ref{sec:baseline}, where the maximisers are well
defined, and involves \(\tilde{J}^n\) and \(\tilde{\zeta}^{n,\cdot}\) only
through their values and first derivatives at the baseline quotes.

\begin{rem}
The value function \(\vartheta\) and the optimal quotes depend on the discount rate \(\rho\) whose value may appear arbitrary. When \(\rho \to 0\), the value function blows up, but \(\rho \vartheta\) converges towards a finite value independent of \(q\) called the ergodic constant of the problem. More importantly, finite differences of the value function, i.e. terms of the form \(D^k_+ \vartheta(q)\) and \(D^k_- \vartheta(q)\), converge, hence the convergence of optimal quotes when \(\rho \to 0\).
\end{rem}

\subsection{The Baseline without Informational Risk}
\label{sec:baseline}

Solving Eq.~\eqref{eq:thetaModel_main} in full generality requires the introduction of assumptions on the functions \( \tilde{J}^{\cdot} \) and \( \tilde{\zeta}^{\cdot,\cdot} \). Since our objective is instead to investigate first-order informational effects, we begin with the classical baseline case where there is no informational risk -- that is, we assume \( \tilde{J}^{n} = 0 \) and \( \tilde{\zeta}^{n,k} = 0 \) for all \( (n,k) \in \mathcal{N} \times \mathcal{K} \). In this setting, the reference price evolves as a Brownian motion and the baseline value function \( \theta : \Delta^1 \mathbb Z \to \mathbb{R} \) solves the simpler HJB equation:
\begin{equation*}
0 =\ -\rho\, \theta(q) - \frac{\gamma}{2}\sigma^2 q^2 + \sum_{\substack{n \in \mathcal{N} \\ k \in \mathcal{K}}}  \Delta^k H^{n,k,b}\left(D^k_+ \theta(q)\right) + \sum_{\substack{n \in \mathcal{N} \\ k \in \mathcal{K}}} \Delta^k H^{n,k,a}\left(D^k_- \theta(q)\right),
\end{equation*}
where the Hamiltonian functions are defined, for all \((n,k) \in \mathcal{N}\times\mathcal{K}\), by
\begin{equation*}
H^{n,k,b/a}(p) := \sup_{\delta \in \mathbb{R}}  \Lambda^{n,k,b/a}(\delta)\, (\delta - p),
\end{equation*}
and the associated maximisers\footnote{They verify the simple ordinary differential equation \(\tilde{\delta}^{n,k,b/a*'}(p) = \frac{1}{c^{n,k,b/a}(\tilde{\delta}^{n,k,b/a*}(p))}\).} are given by
\begin{equation*}
\tilde{\delta}^{n,k,b/a*}(p) := \underset{\delta \in \mathbb{R}}{\text{argmax}}\ \Lambda^{n,k,b/a}(\delta)\, (\delta - p) = \left( \Lambda^{n,k,b/a} \right)^{-1}\left( - {H^{n,k,b/a\prime}}(p) \right).
\end{equation*}

Given the baseline value function \( \theta \), the baseline optimal quotes in feedback form are therefore the following, for all \((n,k) \in \mathcal{N}\times\mathcal{K}\):
\begin{equation*}
\delta^{n,k,b*}(q) = \tilde{\delta}^{n,k,b*}\left(D^k_+ \theta(q)\right)  \quad \text{and} \quad 
\delta^{n,k,a*}(q) = \tilde{\delta}^{n,k,a*}\left(D^k_- \theta(q)\right).
\end{equation*}

\subsection{First-Order Effects}
\label{sec:foe}

\subsubsection{Framework}

In order to study the first-order effects of adverse selection and price reading, we introduce a perturbation parameter \( \varepsilon > 0 \) and set
\[
\tilde{J}^{n}(x) = \varepsilon J^{n}(x) \quad \text{for all } n \in \mathcal{N}\qquad  \text{and} \qquad \tilde{\zeta}^{n,k}(\delta) = \varepsilon \zeta^{n,k}(\delta) \quad \text{for all } (n,k) \in \mathcal{N}\times\mathcal{K}.
\]
For notational convenience, we define the \( \varepsilon \)-dependent Hamiltonian functions by
\[
\mathcal{H}^{n}\left(q, p^{\cdot,b}, p^{\cdot,a}, \varepsilon \right) 
:=\ \tilde{\mathcal{H}}^n\left[ \varepsilon J^n , \varepsilon \zeta^{n,\cdot}  \right]
\left(q, p^{\cdot,b}, p^{\cdot,a} \right) =\ \sup_{ \delta^{n,\cdot,b}, \delta^{n,\cdot,a} } 
\mathcal{B}^n\Big(q, p^{\cdot,b}, p^{\cdot,a}, \delta^{n,\cdot,b}, \delta^{n,\cdot,a}, \varepsilon \Big),
\]
where
\begin{align*}
\mathcal{B}^n\Big(&q, p^{\cdot,b}, p^{\cdot,a}, 
         \delta^{n,\cdot,b}, \delta^{n,\cdot,a}, \varepsilon \Big)
:=\ q\, \varepsilon J^n\left( \sum_{k \in \mathcal{K}} w^{n,k} 
\left( \delta^{n,k,a} - \delta^{n,k,b} \right) \right) \nonumber \\
&+ \sum_{k \in \mathcal{K}} \Delta^k\, \Lambda^{n,k,b}\left( \delta^{n,k,b} \right)
\left( \delta^{n,k,b} - p^{k,b} - \frac{q + \Delta^k}{\Delta^k} \varepsilon \zeta^{n,k}\left( \delta^{n,k,b} \right) \right) \nonumber \\
&+ \sum_{k \in \mathcal{K}} \Delta^k\, \Lambda^{n,k,a}\left( \delta^{n,k,a} \right)
\left( \delta^{n,k,a} - p^{k,a} + \frac{q - \Delta^k}{\Delta^k} \varepsilon \zeta^{n,k}\left( \delta^{n,k,a} \right) \right).
\end{align*}

\subsubsection{Value Function}

Using a formal expansion, we then seek a solution to Eq.~\eqref{eq:thetaModel_main} of the form:
\[
\vartheta(q) = \theta(q) + \varepsilon f(q) + o(\varepsilon),
\]
where \( f \) captures the first-order impact of informational effects on the value function.

To compute the equation for \( f \), we need the gradient of the Hamiltonian functions. Notice that:

\[
\nabla_{\begin{psmallmatrix}
p^{\cdot,b} \\
p^{\cdot,a}
\end{psmallmatrix}} 
\mathcal{H}^n\left(q, p^{\cdot,b}, p^{\cdot,a}, 0 \right)
= - \begin{pmatrix}
\Delta^\cdot\, \Lambda^{n,\cdot,b}\left(\tilde{\delta}^{n,\cdot,b*}(p^{\cdot,b})\right)  \\
\Delta^\cdot\, \Lambda^{n,\cdot,a}\left(\tilde{\delta}^{n,\cdot,a*}(p^{\cdot,a})\right) 
\end{pmatrix} = \begin{pmatrix}
 \Delta^\cdot\, H^{n,\cdot,b}\,'\left(p^{\cdot,b}\right)  \\
\Delta^\cdot\, H^{n,\cdot,a}\,'\left(p^{\cdot,a}\right) 
\end{pmatrix}\]
and
\begin{align*}
\partial_{\varepsilon} \mathcal{H}^{n}(q, p^{\cdot,b}, p^{\cdot,a}, 0) &= q J^n\left(\sum_{k \in \mathcal{K}} w^{n,k}\big(\tilde{\delta}^{n,k,a*}(p^{k,a}) - \tilde{\delta}^{n,k,b*}(p^{k,b})\big)\right) - \sum_{k \in \mathcal{K}} (q+\Delta^k) \left(\Lambda^{n,k,b}\zeta^{n,k}\right)\big(\tilde{\delta}^{n,k,b*}(p^{k,b})\big) \\
&\qquad\qquad \qquad \qquad \qquad \qquad \qquad \qquad\qquad \qquad\quad  + \sum_{k \in \mathcal{K}} (q-\Delta^k) \left(\Lambda^{n,k,a}\zeta^{n,k}\right)\big(\tilde{\delta}^{n,k,a*}(p^{k,a})\big).
\end{align*}

Injecting \( \vartheta(q) = \theta(q) + \varepsilon f(q) + o(\varepsilon) \) into Eq.~\eqref{eq:thetaModel_main} and collecting \( \mathcal{O}(\varepsilon) \) terms yields the linear equation:
\begin{align}
\label{eq:f_linearized}
0 = &-\rho f(q) + \sum_{n \in \mathcal{N}} \Bigg( q J^n\left(\sum_{k \in \mathcal{K}} w^{n,k} (\delta^{n,k,a*}(q) -\delta^{n,k,b*}(q))\right)\\
&-\sum_{k \in \mathcal{K}}  \Lambda^{n,k,b}(\delta^{n,k,b*}(q)) \left(f(q) - f(q + \Delta^k) + (q + \Delta^k) \zeta^{n,k}(\delta^{n,k,b*}(q))\right) \nonumber \\
&- \sum_{k \in \mathcal{K}}\Lambda^{n,k,a}(\delta^{n,k,a*}(q)) \left(f(q) - f(q - \Delta^k) - (q - \Delta^k) \zeta^{n,k}(\delta^{n,k,a*}(q))\right) \Bigg). \nonumber
\end{align}

Using a Feynman-Kac representation, and defining \(\delta^{n,k,b/a*}_t =  \delta^{n,k,b/a*}(q_{t-})\), the solution can be written as a sum over tiers
\[
f(q) = \sum_{n \in \mathcal{N}} f^n(q),
\]
where, for each \(n \in \mathcal{N}\),
\begin{align*}
f^n(q) = \mathbb{E} \Bigg[ \int_0^{\infty} e^{-\rho t} \Bigg(& q_{t-} J^n\Big(\sum_{k \in \mathcal{K}} w^{n,k} (\delta^{n,k,a*}_t - \delta^{n,k,b*}_t)\Big) \, dt \\
&- \sum_{k \in \mathcal{K}} \Big((q_{t-} + \Delta^k) \zeta^{n,k}(\delta^{n,k,b*}_t) \, dN^{n,k,b}_t - (q_{t-} - \Delta^k) \zeta^{n,k}(\delta^{n,k,a*}_t) \, dN^{n,k,a}_t\Big) \Bigg)\,\bigg|\, q_0 = q \Bigg].
\end{align*}

In other words, \(f\) measures the impact of informational risk on the risk-adjusted expected PnL if the market maker kept the quotes that were optimal in the absence of informational risk.

\subsubsection{Optimal Quotes}
\label{foe3}
The above first-order expansion quantifies the impact of informational risk on the risk-adjusted expected PnL of the market maker. In order to address the optimal behaviour of the market maker, we also need a first-order expansion of the maximisers in the definition of the Hamiltonian functions.

Writing
\[
\tilde{\mathfrak{d}}^{n,k,b/a*}(q,p^{\cdot,b},p^{\cdot,a}) = \tilde \delta^{n,k,b/a*}(p^{k,b/a}) + \varepsilon g^{n,k,b/a}(q,p^{\cdot,b},p^{\cdot,a}) + o(\varepsilon),
\]
we have
\begin{align*}
\nabla_{\begin{psmallmatrix}
\delta^{n,\cdot,b}\\
\delta^{n,\cdot,a}
\end{psmallmatrix}} 
\mathcal{B}^n\Big(& q,\, p^{\cdot,b}, 
p^{\cdot,a}, \tilde \delta^{n,\cdot,b*}(p^{\cdot,b}) + \varepsilon g^{n,\cdot,b}(q,p^{\cdot,b},p^{\cdot,a}), 
 \tilde \delta^{n,\cdot,a*}(p^{\cdot,a}) + \varepsilon g^{n,\cdot,a}(q,p^{\cdot,b},p^{\cdot,a}), \varepsilon \Big) = o(\varepsilon).
\end{align*}
Expanding in \(\varepsilon\), and using the first-order condition in the baseline case, we get
\begin{align*}
&\nabla^2_{\begin{psmallmatrix}
\delta^{n,\cdot,b}\\
\delta^{n,\cdot,a}
\end{psmallmatrix}, 
\begin{psmallmatrix}
\delta^{n,\cdot,b}\\
\delta^{n,\cdot,a}
\end{psmallmatrix}} 
\mathcal{B}^n\Big( q,\, p^{\cdot,b},\, p^{\cdot,a},\, 
\tilde\delta^{n,\cdot,b*}(p^{\cdot,b}),\, \tilde\delta^{n,\cdot,a*}(p^{\cdot,a}),\, 0 \Big)
\begin{pmatrix}
g^{n,\cdot,b}(q,p^{\cdot,b},p^{\cdot,a}) \\
g^{n,\cdot,a}(q,p^{\cdot,b},p^{\cdot,a})
\end{pmatrix} \\[0.8ex]
+\
&\partial_\varepsilon \nabla_{\begin{psmallmatrix}
\delta^{n,\cdot,b}\\
\delta^{n,\cdot,a}
\end{psmallmatrix}} 
\mathcal{B}^n\Big( q,\, p^{\cdot,b},\, p^{\cdot,a},\, 
\tilde\delta^{n,\cdot,b*}(p^{\cdot,b}),\, \tilde\delta^{n,\cdot,a*}(p^{\cdot,a}),\, 0  \Big)
= 0.
\end{align*}

Solving this linear system yields
\begin{equation*}
g^{n,k,b}(q,p^{\cdot,b},p^{\cdot,a}) = \frac{q w^{n,k} J^{n\,\prime}\!\left( \sum_{j \in \mathcal{K}} w^{n,j} \left( \tilde\delta^{n,j,a*}(p^{j,a}) - \tilde\delta^{n,j,b*}(p^{j,b}) \right) \right)
+ (q+\Delta^k) \left( \Lambda^{n,k,b} \zeta^{n,k} \right)'\!\left( \tilde\delta^{n,k,b*}(p^{k,b}) \right)
}{
\Delta^k\Lambda^{n,k,b\,\prime}\!\left( \tilde\delta^{n,k,b*}(p^{k,b}) \right) 
c^{n,k,b}(\tilde\delta^{n,k,b*}(p^{k,b}))
}
\end{equation*}
and
\begin{equation*}
g^{n,k,a}(q,p^{\cdot,b},p^{\cdot,a}) = \frac{
- q w^{n,k} J^{n\,\prime}\!\left( \sum_{j \in \mathcal{K}} w^{n,j} \left( \tilde\delta^{n,j,a*}(p^{j,a}) - \tilde\delta^{n,j,b*}(p^{j,b}) \right) \right)
- (q-\Delta^k) \left( \Lambda^{n,k,a} \zeta^{n,k} \right)'\!\left( \tilde\delta^{n,k,a*}(p^{k,a}) \right)
}{
\Delta^k\Lambda^{n,k,a\,\prime}\!\left( \tilde\delta^{n,k,a*}(p^{k,a}) \right) 
c^{n,k,a}(\tilde\delta^{n,k,a*}(p^{k,a}))
}.
\end{equation*}

To get the optimal quotes in feedback form, we combine the above two first-order expansions to get

\begin{align*}
\mathfrak{d}^{n,k,b*}(q) &= \tilde{\mathfrak{d}}^{n,k,b*}(q,D^\cdot_+ \vartheta(q),D^\cdot_- \vartheta(q))\\
&=\tilde\delta^{n,k,b*}(D^k_+ \theta(q) + \varepsilon D^k_+ f(q)) + \varepsilon g^{n,k,b}(q,D^\cdot_+ \theta(q) + \varepsilon D^\cdot_+ f(q),D^\cdot_- \theta(q)+ \varepsilon D^\cdot_- f(q)) + o(\varepsilon)\\
&= \tilde\delta^{n,k,b*}(D^k_+ \theta(q)) + \varepsilon \tilde\delta^{n,k,b*'}(D^k_+ \theta(q)) D^k_+ f(q) + \varepsilon g^{n,k,b}(q,D^\cdot_+ \theta(q),D^\cdot_- \theta(q)) + o(\varepsilon) \\
&= \delta^{n,k,b*}(q) + \frac{\varepsilon}{c^{n,k,b}(\delta^{n,k,b*}(q))}  \Bigg(D^k_+ f(q) + \frac{qw^{n,k} J^{n\,\prime}\!\left( \sum_{j \in \mathcal{K}} w^{n,j} \left( \delta^{n,j,a*}(q) - \delta^{n,j,b*}(q) \right) \right)}{{\Delta^k}  \Lambda^{n,k,b\,\prime}\!\left(\delta^{n,k,b*}(q) \right)}\\
& \qquad\qquad\qquad\qquad\qquad\qquad\qquad\qquad + \frac{(q+\Delta^k) \left( \Lambda^{n,k,b} \zeta^{n,k} \right)'\!\left(\delta^{n,k,b*}(q) \right)
}{
{\Delta^k} \Lambda^{n,k,b\,\prime}\!\left(\delta^{n,k,b*}(q) \right) 
}\Bigg) + o(\varepsilon)
\end{align*}
and similarly
\begin{align*}
\mathfrak{d}^{n,k,a*}(q) &= \delta^{n,k,a*}(q) + \frac{\varepsilon}{c^{n,k,a}(\delta^{n,k,a*}(q))}  \Bigg(D^k_- f(q) - \frac{q w^{n,k} J^{n\,\prime}\!\left( \sum_{j \in \mathcal{K}} w^{n,j} \left( \delta^{n,j,a*}(q) - \delta^{n,j,b*}(q) \right) \right)}{{\Delta^k} \Lambda^{n,k,a\,\prime}\!\left(\delta^{n,k,a*}(q) \right)}\\
& \qquad\qquad\qquad\qquad\qquad\qquad\qquad\qquad - \frac{(q-\Delta^k) \left( \Lambda^{n,k,a} \zeta^{n,k} \right)'\!\left(\delta^{n,k,a*}(q) \right)
}{
{\Delta^k} \Lambda^{n,k,a\,\prime}\!\left(\delta^{n,k,a*}(q) \right) 
}\Bigg) + o(\varepsilon).
\end{align*}

These expressions highlight that the influence of informational risk on optimal quotes can be decomposed into two components. The first component arises from the terms \( D^{k}_{\pm} f \), which are finite differences of the function \( f \). By definition, these terms reflect the marginal impact of informational risk on the risk-adjusted expected PnL, assuming the market maker does not update their quotes. As such, they aggregate the effects of informational risk across all tiers and are identical for all \(n \in \mathcal N\). The second component is represented by the terms \( g^{n,k,b/a} \), which quantify the impact of informational risk on the quotes derived from the baseline value function in the absence of informational risk. Unlike the \( D^{k}_{\pm} f \) terms, the \( g^{n,k,b/a} \) terms are specific to the considered tier.

These two components may push optimal quotes in opposite directions, and their
net effect typically requires numerical evaluation. Numerical computation raises
no difficulty, as it relies only on the optimal quotes in the baseline setting
and on the function \( f \), which solves the infinite linear system
\eqref{eq:f_linearized} and can be approximated on a truncated inventory grid,
large enough for the results to be stable on the inventory range of interest. We
nevertheless prefer, in what follows, to rely on a quadratic approximation of
the Hamiltonian functions: it yields closed-form approximations for \( f \) and for the optimal quotes, and thereby sheds light on the mechanisms at play.

\section{Quadratic Hamiltonian Approximation}
\label{sec:begv}

In the algorithmic market making literature, closed-form approximations of the value function and optimal quotes are frequently used. One approach (see \cite{bergault2021closed}) consists in replacing the HJB equation associated with the control problem by an alternative HJB equation in which the Hamiltonian functions are approximated to second order. The resulting modified HJB equation admits a solution that is quadratic in the inventory variable, with coefficients computable through a Riccati equation\footnote{In our stationary case, the algebraic Riccati equation reduces to a simple quadratic equation.} and a linear equation.\footnote{While we focus on the single-asset case in this article, it is worth emphasising that this methodology extends naturally to multi-asset market making, thereby offering an effective remedy to the curse of dimensionality.}

\subsection{Towards Quadratic Hamiltonian Functions}

In this section, we expand the Hamiltonian functions to second order with respect to \( \left( p^{\cdot,b}, p^{\cdot,a}, \varepsilon \right) \) around the point~\( (0,0,0) \).

From Section~\ref{sec:foe}, we have the following first-order derivatives:
\begin{align*}
\nabla_{\begin{psmallmatrix}
p^{\cdot,b} \\
p^{\cdot,a}
\end{psmallmatrix}} 
\mathcal{H}^n\left(q, 0, 0, 0 \right)
&= \begin{pmatrix}
  \Delta^\cdot\, H^{n,\cdot,b}\,'(0)  \\
  \Delta^\cdot\, H^{n,\cdot,a}\,'(0) 
\end{pmatrix},
\end{align*}
and
\begin{align*}
\partial_{\varepsilon} \mathcal{H}^{n}(q, 0, 0, 0) &= q\, J^n\left(\sum_{k \in \mathcal{K}} w^{n,k}\left(\tilde{\delta}^{n,k,a*}(0) - \tilde{\delta}^{n,k,b*}(0)\right)\right) - \sum_{k \in \mathcal{K}} (q+\Delta^k)\, \left(\Lambda^{n,k,b}\zeta^{n,k}\right)\left(\tilde{\delta}^{n,k,b*}(0)\right) \\
& \qquad\qquad\qquad\qquad\qquad\qquad\qquad\qquad\qquad\quad+ \sum_{k \in \mathcal{K}}  (q-\Delta^k)\, \left(\Lambda^{n,k,a}\zeta^{n,k}\right)\left(\tilde{\delta}^{n,k,a*}(0)\right)\, .
\end{align*}

\begin{rem}
For all \( (n,k) \in \mathcal{N}\times\mathcal{K} \), the quotes \( \tilde{\delta}^{n,k,b*}(0) \) and \( \tilde{\delta}^{n,k,a*}(0) \) correspond to the myopic quotes that a market maker with no risk aversion would set in the baseline case without informational risk.
\end{rem}
For the second-order derivatives, we have
\begin{align*}
&\nabla^2_{\begin{psmallmatrix}
p^{\cdot,b} \\
p^{\cdot,a} \\
\varepsilon
\end{psmallmatrix},
\begin{psmallmatrix}
p^{\cdot,b} \\
p^{\cdot,a} \\
\varepsilon
\end{psmallmatrix}}
\mathcal{H}^n\left(q, 0, 0, 0 \right)\\
= &\nabla^2_{\begin{psmallmatrix}
p^{\cdot,b} \\
p^{\cdot,a} \\
\varepsilon
\end{psmallmatrix},
\begin{psmallmatrix}
p^{\cdot,b} \\
p^{\cdot,a} \\
\varepsilon
\end{psmallmatrix}}
\mathcal{B}^n(\bar{x}^n) - \nabla^2_{\begin{psmallmatrix}
p^{\cdot,b} \\
p^{\cdot,a} \\
\varepsilon
\end{psmallmatrix},
\begin{psmallmatrix}
\delta^{n,\cdot,b} \\
\delta^{n,\cdot,a}
\end{psmallmatrix}}
\mathcal{B}^n(\bar{x}^n)
\left( \nabla^2_{\begin{psmallmatrix}
\delta^{n,\cdot,b} \\
\delta^{n,\cdot,a}
\end{psmallmatrix},
\begin{psmallmatrix}
\delta^{n,\cdot,b} \\
\delta^{n,\cdot,a}
\end{psmallmatrix}}
\mathcal{B}^n(\bar{x}^n) \right)^{-1} \nabla^2_{\begin{psmallmatrix}
\delta^{n,\cdot,b} \\
\delta^{n,\cdot,a}
\end{psmallmatrix},
\begin{psmallmatrix}
p^{\cdot,b} \\
p^{\cdot,a} \\
\varepsilon
\end{psmallmatrix}}
\mathcal{B}^n(\bar{x}^n),
\end{align*}
where \[
\bar{x}^n := \left(q,\, 0,\, 0,\, \tilde{\delta}^{n,\cdot,b*}(0),\, \tilde{\delta}^{n,\cdot,a*}(0),\, 0 \right).
\]

Now, we easily see that
\[
\nabla^2_{\begin{psmallmatrix}
p^{\cdot,b} \\
p^{\cdot,a} \\
\varepsilon
\end{psmallmatrix}, \begin{psmallmatrix}
p^{\cdot,b} \\
p^{\cdot,a} \\
\varepsilon
\end{psmallmatrix}} 
\mathcal{B}^n\left(\bar{x}^n\right) = 0,
\]
and that
\[
\nabla^2_{\begin{psmallmatrix}
\delta^{n,\cdot,b} \\
\delta^{n,\cdot,a}
\end{psmallmatrix}, \begin{psmallmatrix}
\delta^{n,\cdot,b} \\
\delta^{n,\cdot,a}
\end{psmallmatrix}} 
\mathcal{B}^n\left(\bar{x}^n \right)
=
\operatorname{diag} \left(
 -\Delta^\cdot \dfrac{\Lambda^{n,\cdot,b\,\prime}(\tilde\delta^{n,\cdot,b*}(0))^2 }{ H^{n,\cdot,b\,\prime\prime}(0) } ,
-\Delta^\cdot \dfrac{ \Lambda^{n,\cdot,a\,\prime}(\tilde\delta^{n,\cdot,a*}(0))^2 }{ H^{n,\cdot,a\,\prime\prime}(0) } 
\right),
\]
while
\begin{align*}
\nabla^2_{\begin{psmallmatrix}
p^{\cdot,b} \\
p^{\cdot,a} \\
\varepsilon
\end{psmallmatrix}, 
\begin{psmallmatrix}
\delta^{n,\cdot,b} \\
\delta^{n,\cdot,a}
\end{psmallmatrix}} 
\mathcal{B}^n\left(\bar{x}^n \right)
&=
\begin{pmatrix}
\operatorname{diag}\left(-\Delta^\cdot\, \Lambda^{n,\cdot,b\,\prime}\!\left(\tilde{\delta}^{n,\cdot,b*}(0)\right)\right) & 0 \\[1ex]
0 & \operatorname{diag}\left(-\Delta^\cdot\, \Lambda^{n,\cdot,a\,\prime}\!\left(\tilde{\delta}^{n,\cdot,a*}(0)\right)\right) \\[1ex]
\eta^{\cdot \top} & \iota^{\cdot \top}
\end{pmatrix},
\end{align*}
where, for all \(k \in \mathcal K\),
\[
\eta^k := -q\, w^{n,k}\, J^{n\,\prime}\!\left(\sum_{j \in \mathcal{K}} w^{n,j} \left(\tilde{\delta}^{n,j,a*}(0) - \tilde{\delta}^{n,j,b*}(0)\right) \right) 
- (q + \Delta^k)\, \left(\Lambda^{n,k,b} \zeta^{n,k}\right)'\!\left( \tilde{\delta}^{n,k,b*}(0) \right)
\]
and
\[
\iota^k := \phantom{-}q\, w^{n,k}\, J^{n\,\prime}\!\left(\sum_{j \in \mathcal{K}} w^{n,j} \left(\tilde{\delta}^{n,j,a*}(0) - \tilde{\delta}^{n,j,b*}(0)\right) \right) 
+ (q - \Delta^k)\, \left(\Lambda^{n,k,a} \zeta^{n,k}\right)'\!\left( \tilde{\delta}^{n,k,a*}(0) \right).
\]
Therefore, 

\[
\nabla^2_{\begin{psmallmatrix}
p^{\cdot,b} \\
p^{\cdot,a} \\
\varepsilon
\end{psmallmatrix},
\begin{psmallmatrix}
p^{\cdot,b} \\
p^{\cdot,a} \\
\varepsilon
\end{psmallmatrix}}
\mathcal{H}^n\left(q, 0, 0, 0 \right)
\]
\[
=
\begin{pmatrix}
\Delta^\cdot\, H^{n,\cdot,b\,\prime\prime}(0) & 0 &  -\dfrac{H^{n,\cdot,b\,\prime\prime}(0)}{ \Lambda^{n,\cdot,b\,\prime}\left(\tilde{\delta}^{n,\cdot,b*}(0)\right)}\, \eta^\cdot  \\[1.5ex]
0 & \Delta^\cdot\, H^{n,\cdot,a\,\prime\prime}(0) &  -\dfrac{H^{n,\cdot,a\,\prime\prime}(0)}{ \Lambda^{n,\cdot,a\,\prime}\left(\tilde{\delta}^{n,\cdot,a*}(0)\right)}\, \iota^\cdot  \\[1.5ex]
-\dfrac{H^{n,\cdot,b\,\prime\prime}(0)}{ \Lambda^{n,\cdot,b\,\prime}\left(\tilde{\delta}^{n,\cdot,b*}(0)\right)}\, \eta^{\cdot \top} &  -\dfrac{H^{n,\cdot,a\,\prime\prime}(0)}{ \Lambda^{n,\cdot,a\,\prime}\left(\tilde{\delta}^{n,\cdot,a*}(0)\right)}\, \iota^{\cdot \top} &
\begin{aligned}
\sum_{k \in \mathcal{K}} & \left[
\dfrac{H^{n,k,b\,\prime\prime}(0)}{ \Delta^k\, \Lambda^{n,k,b\,\prime}\left(\tilde{\delta}^{n,k,b*}(0)\right)^2 }\, {\eta^k}^2 \right. \\
&\left. + \dfrac{H^{n,k,a\,\prime\prime}(0)}{ \Delta^k\, \Lambda^{n,k,a\,\prime}\left(\tilde{\delta}^{n,k,a*}(0)\right)^2 }\, {\iota^k}^2
\right]
\end{aligned}
\end{pmatrix}.
\]

\subsection{HJB Equation and Value Function}

We now consider, instead of Eq.~\eqref{eq:thetaModel_main}, the following equation:
\begin{align}
\label{eq:thetaquad}
0 =\ & -\rho\, \widehat\vartheta(q) - \frac{\gamma}{2} \sigma^2 q^2
+ \sum_{n \in \mathcal{N}} \mathcal{H}^n\left(q, 0, 0, 0 \right) + 
\sum_{n \in \mathcal{N}}
\nabla_{\begin{psmallmatrix}
p^{\cdot,b} \\
p^{\cdot,a} \\
\varepsilon
\end{psmallmatrix}} 
\mathcal{H}^n\left(q, 0, 0, 0 \right)^\top
\begin{pmatrix}
D^\cdot_+ \widehat\vartheta(q) \\
D^\cdot_- \widehat\vartheta(q) \\
\varepsilon
\end{pmatrix} \nonumber \\
&+ \frac{1}{2}
\sum_{n \in \mathcal{N}}
\begin{pmatrix}
D^\cdot_+ \widehat\vartheta(q) \\
D^\cdot_- \widehat\vartheta(q) \\
\varepsilon
\end{pmatrix}^\top
\nabla^2_{\begin{psmallmatrix}
p^{\cdot,b} \\
p^{\cdot,a} \\
\varepsilon
\end{psmallmatrix}, \begin{psmallmatrix}
p^{\cdot,b} \\
p^{\cdot,a} \\
\varepsilon
\end{psmallmatrix}}
\mathcal{H}^n\left(q, 0, 0, 0 \right)
\begin{pmatrix}
D^\cdot_+ \widehat\vartheta(q) \\
D^\cdot_- \widehat\vartheta(q) \\
\varepsilon
\end{pmatrix}.
\end{align}

A straightforward computation shows that the solution \(\widehat \vartheta\) to Eq. \eqref{eq:thetaquad} is of the form \(\widehat \vartheta(q) = - A_{\varepsilon} q^2 - B_{\varepsilon} q - C_{\varepsilon}\) where \(A_{\varepsilon} > 0\) and \(B_{\varepsilon}\) satisfy\footnote{We do not report any equation for \(C_{\varepsilon}\) as it does not play any role for the optimal quotes. Note that, when \(\rho \to 0\), \(C_\varepsilon\) typically blows up, but \(\rho C_\varepsilon\) converges towards a finite value.} the following equations: 
\begin{equation}
\label{eq:A}
\rho A_{\varepsilon} - \frac{\gamma}{2} \sigma^2 + 2 A_{\varepsilon}^2 F^{2,1}_+ + 2 \varepsilon A_{\varepsilon} \Sigma_+^{1,0} = \mathcal{O}(\varepsilon^2),
\end{equation}
\begin{equation}
\label{eq:B}
\rho B_{\varepsilon} 
+ 2A_{\varepsilon}B_{\varepsilon} F^{2,1}_+ + 2A_{\varepsilon}^2 F^{2,2}_- +  2A_{\varepsilon} F^{1,1}_- + \varepsilon B_{\varepsilon} \Sigma^{1,0}_+ + \varepsilon A_{\varepsilon} \Sigma^{3,1}_- + \varepsilon \Xi = \mathcal{O}(\varepsilon^2),
\end{equation}
where
\[F^{i,l}_{\pm} = \sum_{\substack{n \in \mathcal{N} \\ k \in \mathcal{K}}} \left(\Delta^k\right)^l \left( {H^{n,k,b}}^{(i)}(0) \pm {H^{n,k,a}}^{(i)}(0) \right),\]
\begin{align*}
\Sigma^{r,l}_{\pm} &=  \sum_{\substack{n \in \mathcal{N} \\ k \in \mathcal{K}}} \left(\Delta^k\right)^l \left[
\frac{H^{n,k,b\,\prime\prime}(0)}{\Lambda^{n,k,b\,\prime}\left(\tilde{\delta}^{n,k,b*}(0)\right)} 
\left( w^{n,k} J^{n\,\prime}\!\left( \sum_{j \in \mathcal{K}} w^{n,j} (\tilde{\delta}^{n,j,a*}(0) - \tilde{\delta}^{n,j,b*}(0)) \right) 
+ r \left( \Lambda^{n,k,b} \zeta^{n,k} \right)'\!\left(\tilde{\delta}^{n,k,b*}(0)\right) \right) \right. \\
&\left. \quad \pm 
\frac{H^{n,k,a\,\prime\prime}(0)}{\Lambda^{n,k,a\,\prime}\left(\tilde{\delta}^{n,k,a*}(0)\right)} 
\left( w^{n,k} J^{n\,\prime}\!\left( \sum_{j \in \mathcal{K}} w^{n,j} (\tilde{\delta}^{n,j,a*}(0) - \tilde{\delta}^{n,j,b*}(0)) \right) 
+ r \left( \Lambda^{n,k,a} \zeta^{n,k} \right)'\!\left(\tilde{\delta}^{n,k,a*}(0)\right) \right) 
\right]
\end{align*}
and
\[\Xi = \sum_{n \in \mathcal{N}} \left[
J^n\left( \sum_{j \in \mathcal{K}} w^{n,j} \left( \tilde{\delta}^{n,j,a*}(0) - \tilde{\delta}^{n,j,b*}(0) \right) \right)
- \sum_{k \in \mathcal{K}} \left(
\left( \Lambda^{n,k,b} \zeta^{n,k} \right)\!\left( \tilde{\delta}^{n,k,b*}(0) \right) 
- \left( \Lambda^{n,k,a} \zeta^{n,k} \right)\!\left( \tilde{\delta}^{n,k,a*}(0) \right)
\right) \right].\]

From Eqs. \eqref{eq:A} and \eqref{eq:B}, we get
\[ A_\varepsilon = A_0 + \varepsilon A_0' + o(\varepsilon) \qquad \text{and} \qquad B_\varepsilon = B_0 + \varepsilon B_0' + o(\varepsilon),\]
where
\[\left\{
\begin{aligned}
A_0 &= \frac{-\rho + \sqrt{\rho^2 + 4\gamma \sigma^2 F_+^{2,1} }}{4F_+^{2,1}} \\
B_0 &= -\frac{2A_0}{\rho + 2A_0 F_+^{2,1}}\left(A_0 F_-^{2,2} + F_-^{1,1}\right) \\
A_0' &= -\frac{2A_0\Sigma_+^{1,0}}{\rho + 4A_0 F_+^{2,1}} \\
B_0' &= -\frac{1}{\rho + 2A_0 F_+^{2,1}}\left( B_0 \Sigma^{1,0}_+ + A_0 \Sigma^{3,1}_- + \Xi - \frac{4A_0\Sigma_+^{1,0}}{\rho + 4A_0 F_+^{2,1}}\left(B_0 F^{2,1}_+ + 2 A_0 F^{2,2}_- + F^{1,1}_-\right) \right).
\end{aligned}
\right.\]

Therefore, if \(\widehat{\vartheta}_0(q) = - A_0 q^2 - B_0 q\) and  \(\widehat f(q) = -A_0' q^2 - B_0'q\), then our approximation of the value function using quadratic approximations of Hamiltonian functions is (up to an additive constant) given by:
\[\widehat{\vartheta}(q) = \widehat{\vartheta}_0(q) + \varepsilon \widehat f(q)+o(\varepsilon).\]

\subsection{Optimal Quotes}

If the quotes in feedback form associated with \(\widehat{\vartheta}_0\), i.e. in the case of no informational risk under the quadratic approximation, are denoted by
\begin{equation}
\label{num_quotes_no}
\widehat\delta^{n,k,b*}(q) := \tilde{\delta}^{n,k,b*}\left(D^k_+ \widehat\vartheta_0(q)\right)  \quad \text{and} \quad 
\widehat\delta^{n,k,a*}(q) := \tilde{\delta}^{n,k,a*}\left(D^k_- \widehat\vartheta_0(q)\right),
\end{equation}
then, as in Section \ref{foe3}, the first-order effects of informational risk on optimal quotes in feedback form are given by
\begin{align}
\label{num_quotes_1}
\widehat{\mathfrak{d}}^{n,k,b*}(q) &= \widehat\delta^{n,k,b*}(q) + \frac{\varepsilon}{c^{n,k,b}\left(\widehat\delta^{n,k,b*}(q)\right)}  \left(D^k_+ \widehat f(q) + \frac{qw^{n,k} J^{n\,\prime}\!\left( \sum_{j \in \mathcal{K}} w^{n,j} \left( \widehat\delta^{n,j,a*}(q) - \widehat\delta^{n,j,b*}(q) \right) \right)}{{\Delta^k}  \Lambda^{n,k,b\,\prime}\!\left(\widehat\delta^{n,k,b*}(q) \right)}\right.\nonumber\\
& \qquad\qquad\qquad\qquad\qquad\qquad\qquad\qquad \left.+ \frac{(q+\Delta^k) \left( \Lambda^{n,k,b} \zeta^{n,k} \right)'\!\left(\widehat\delta^{n,k,b*}(q) \right)
}{
{\Delta^k} \Lambda^{n,k,b\,\prime}\!\left(\widehat\delta^{n,k,b*}(q) \right) 
}\right) + o(\varepsilon)
\end{align}
and
\begin{align}
\label{num_quotes_2}
\widehat{\mathfrak{d}}^{n,k,a*}(q) &= \widehat\delta^{n,k,a*}(q) + \frac{\varepsilon}{c^{n,k,a}\left(\widehat\delta^{n,k,a*}(q)\right)}  \left(D^k_- \widehat f(q) - \frac{qw^{n,k} J^{n\,\prime}\!\left( \sum_{j \in \mathcal{K}} w^{n,j} \left( \widehat\delta^{n,j,a*}(q) - \widehat\delta^{n,j,b*}(q) \right) \right)}{{\Delta^k}  \Lambda^{n,k,a\,\prime}\!\left(\widehat\delta^{n,k,a*}(q) \right)}\right.\nonumber\\
& \qquad\qquad\qquad\qquad\qquad\qquad\qquad\qquad \left.- \frac{(q-\Delta^k) \left( \Lambda^{n,k,a} \zeta^{n,k} \right)'\!\left(\widehat\delta^{n,k,a*}(q) \right)
}{
{\Delta^k} \Lambda^{n,k,a\,\prime}\!\left(\widehat\delta^{n,k,a*}(q) \right) 
}\right) + o(\varepsilon).
\end{align}

\subsection{Analysis in the Symmetric Case}

Given that \(\widehat{f}\) is available in closed form, the above expressions are easier to handle than those in Section~\ref{foe3}. This allows us to examine the underlying mechanisms in detail. For simplicity, we focus on a symmetric setting with identical bid- and ask-side functions, in the limiting case where \(\rho \to 0\).

In this case indeed, we have \(B_0 = B'_0 = 0\) because of the symmetry, and the first-order adjustment to the value function is, up to an additive constant, given by
\[ \widehat{f}(q) = \frac 12\frac{\Sigma_+^{1,0}}{F_+^{2,1}}\, q^2, \] where\footnote{We removed the \(b\) and \(a\) superscripts because of our assumption of symmetry.}
\[
\Sigma^{1,0}_{+} =  2\sum_{\substack{m \in \mathcal{N} \\ j \in \mathcal{K}}}
\frac{H^{m,j\,\prime\prime}(0)}{\Lambda^{m,j\,\prime}\left(\tilde{\delta}^{m,j*}(0)\right)} 
\left( w^{m,j} J^{m\,\prime}\!\left(0\right) 
+ \left( \Lambda^{m,j} \zeta^{m,j} \right)' \left(\tilde{\delta}^{m,j*}(0)\right) \right) \quad \text{and} \quad F^{2,1}_{+} = 2\sum_{\substack{m \in \mathcal{N} \\ j \in \mathcal{K}}} \Delta^j H^{m,j\,\prime\prime}(0).\]

Because \( J^n \) is nondecreasing, price reading contributes to increasing the concavity of the value function. The intuition is as follows: in the absence of informational risk, the market maker skews quotes to attract offsetting flows and deter trades that would increase the absolute value of the inventory. Skew sniffers detect this behaviour and push the price in a direction unfavourable to the market maker, thereby amplifying inventory risk.

In contrast, the impact of adverse selection on the value function requires a more nuanced discussion. Each tier \( n \in \mathcal{N} \) and size index \( k \in \mathcal{K} \) contributes either positively or negatively to the change in concavity of the value function, depending on the sign of \(
\left( \zeta^{n,k} \Lambda^{n,k} \right)'\). The sign of this term is determined by the nature of adverse selection within each tier.

Let us first consider a tier of clients who trade on strong signals and execute in the direction of their signal at virtually any reasonable price given prevailing market conditions. Such behaviour is plausible for clients acting on mid- to long-term signals of the order of several dozen basis points to a few percentage points. Indeed, over the relevant quote range, variations in the offset are small relative to the expected price move, resulting in a mild dependence of \(\zeta\) on the quote. In this case, the slopes of the corresponding \( \zeta \) functions are insufficiently steep to compensate for the decrease of intensity functions, resulting in a reduction in the concavity of the value function. The mechanism is as follows: if the market maker holds a positive (resp. negative) inventory, then by skewing -- through a decrease (resp. increase) in both bid and ask prices -- they raise the absolute and relative probability of executing a sell (resp. buy) order that moves the inventory closer to zero and therefore increase the likelihood that adverse selection -- whose dependence on quotes is mild -- induces a price increase (resp. decrease), thereby increasing the value of the remaining inventory. Although adverse selection reduces the constant term in the approximation of the value function, this mechanism tends to flatten its shape by lowering its concavity.

Conversely, consider a tier where adverse selection predominantly arises when clients execute trades acting on short-term signals of a few basis points thanks to a technological speed advantage. 
Such clients typically trade only when the dealer's quote is sufficiently stale or mispriced, so execution at an apparently unfavourable offset is stronger evidence of a large latent mispricing, leading to a steeper \(\zeta\). In such a scenario, the corresponding \( \zeta \) functions may exhibit steep slopes, and the products \( \zeta^{n,k} \Lambda^{n,k} \) may cease to be decreasing in the neighbourhood of (approximate) baseline quotes. This contributes to an increase in the concavity of the value function.

This analysis of the concavity of the value function provides valuable insight into the first component -- common to all tiers -- describing the impact of informational risk on optimal quotes. This component is proportional to the terms
\[
D^k_{\pm}\widehat{f}(q) = -\frac{\Sigma_+^{1,0}}{F_+^{2,1}}\left(\pm q + \frac {\Delta^k}2 \right).
\]

When the inventory is zero, the preceding discussion justifies that this component implies spread widening in the presence of price reading: the market maker compensates for the added informational risk by increasing the profit margin per trade. When the inventory is nonzero, the effect of price reading on the value function encourages the market maker to skew more aggressively, in order to shorten the time spent with a long or short position. The case of adverse selection is more nuanced, as it depends on the cumulative effect of the different tiers and sizes.

The second component of the optimal quotes, which is tier-specific and relates to the local adjustments of quotes, consists of two terms: one related to price reading and the other to adverse selection.

Focusing first on price reading, the corresponding term in the quote offset has the sign of \(-q\) on the bid side and \(+q\) on the ask side. In particular, it vanishes when the inventory is zero and contributes to reducing the existing skew when the inventory is nonzero. This effect reflects the market maker's incentive to reveal less information through their quotes in order to mitigate the impact of skew sniffers.

Taken together, the two components of the impact of price reading on optimal quotes lead to a subtle trade-off. When the inventory is zero, the second component vanishes and only the first component remains, leading to a widening of the bid-ask spread to compensate for informational risk. When the inventory is nonzero, the first component encourages stronger skewing to reduce the time spent in unbalanced positions, while the second component pushes in the opposite direction, urging the market maker to reduce the skew in order to avoid revealing inventory information to skew sniffers. The second effect dominates, for instance, in tiers with a large proportion of skew sniffers that are highly sensitive to revealed information but trade infrequently.

The case of adverse selection once again depends on the relative shape of the \(\zeta\) and \(\Lambda\) functions. When adverse selection depends only mildly on the quotes, the second component affects the quote offsets according to the sign of \(q + \Delta^k\) for bid transactions of size \(\Delta^k\) and \(-q + \Delta^k\) for ask transactions of the same size. In such cases, the \(\Delta^k\) terms contribute to widening the spread, while the \(q\) terms encourage additional skewing -- opposite to the incentives from the first component. The reasoning is reversed when adverse selection primarily occurs at seemingly unfavourable prices, that is, when informed clients exploit short-term mispricings. However, the second component still pushes in the opposite direction to the first. In both scenarios, the second component may dominate the first in tiers with highly informed clients, especially when trades have a strong impact on prices. However, unlike in the case of price reading, liquidity -- through the amplitude of \(\Lambda^{n,k}\) -- does not play a clear role in this mechanism.

\section{Numerical Examples}
\label{sec:numerics}

\subsection{Functional Forms}
We now provide illustrations in the specific case\footnote{Throughout this section we place ourselves in the symmetric setting and in the limit  \(\rho \to 0\).}
\begin{align}\label{eq:expo_func}
    \Lambda^{n,k}(\delta):= \Lambda^{n,k,b}(\delta) = \Lambda^{n,k,a}(\delta) = \Lambda_0^{n,k} e^{-\kappa^{n,k} \delta}, \qquad 
    J^n(x) = x \quad \text{and} \quad
    \zeta^{n,k}(\delta) = \alpha^{n,k} e^{\beta^{n,k} \delta},
\end{align}
where for all \(n \in \mathcal{N}\) and \(k \in \mathcal{K}\), the parameters \(\Lambda_0^{n,k}, \kappa^{n,k}, \alpha^{n,k}, \beta^{n,k}\) are positive.\footnote{In line with Section~\ref{sec:genpb}, the specifications
\eqref{eq:expo_func} are local: they describe the price-reading response and
the markout profile over the range of quotes relevant around the baseline,
not globally. The linear choice \(J^n(x)=x\) is convenient because our formulas
involve \(J^n\) only through \(J^n(0)=0\) and \(J^{n\,\prime}(0)\), the slope being
understood as \(1\)~s\(^{-1}\) with dimensionless weights \(w^{n,k}\); likewise,
\(\zeta^{n,k}\) enters only through its value and first derivative at the
baseline quotes. Working with \(\mathcal{H}^n\) directly, rather than with
first-order approximations, would call for globally bounded specifications
such as the competitiveness-based variant of the Appendix.} For such exponential intensities, we have
\[c^{n,k,b/a} = 1, \qquad \tilde{\delta}^{n,k*}(p) = p + \frac 1{\kappa^{n,k}} \quad \text{and} \quad H^{n,k}(p) = \frac{\Lambda_0^{n,k}}{\kappa^{n,k}} e^{-1-\kappa^{n,k}p},\] 
so that, in particular, \(H^{n,k\, \prime\prime}(0) = \Lambda_0^{n,k} \kappa^{n,k} e^{-1}\).

Baseline pricing without informational risk in the quadratic approximation case is therefore given by 
\[
\widehat\delta^{n,k,b/a*}(q) = \frac{1}{\kappa^{n,k}} + \sigma \sqrt{\frac{\gamma e}{2\sum_{\substack{m \in \mathcal{N} \\ j \in \mathcal{K}}} \Delta^j  \Lambda_0^{m,j} \kappa^{m,j} }}\left(\pm q + \frac{\Delta^k}{2}\right).
\]

\subsection{Price Reading}
\label{sec:price_reading}

Let us first consider the case of price reading alone.\footnote{We consider price reading and adverse selection separately in order to disentangle their respective effects. 
The first-order corrections are additive in the two channels; cross-effects appear only at second order.} We have
\[
D^k_{\pm}\widehat{f}(q) = \frac{e\sum_{\substack{m \in \mathcal{N} \\ j \in \mathcal{K}}} w^{m,j}}{\sum_{\substack{m \in \mathcal{N} \\ j \in \mathcal{K}}} \Delta^j  \Lambda_0^{m,j} \kappa^{m,j}}\left(\pm q + \frac {\Delta^k}2 \right),
\]
and therefore the quote adjustment is given by
\[\frac{\widehat{\mathfrak{d}}^{n,k,b/a*}(q) - \widehat{\delta}^{n,k,b/a*}(q)}{\varepsilon} = 
	\frac{e\sum_{\substack{m \in \mathcal{N} \\ j \in \mathcal{K}}} w^{m,j}}{\sum_{\substack{m \in \mathcal{N} \\ j \in \mathcal{K}}} \Delta^j  \Lambda_0^{m,j} \kappa^{m,j}} \left(\pm q + \frac{\Delta^k}{2}\right)
	\mp \frac{qw^{n,k}}{\Delta^k \kappa^{n,k}} \frac{1}{\Lambda^{n,k}\left( \widehat\delta^{n,k,b/a*}(q) \right)}.
\]

As discussed earlier, the first term captures the global response to the price reading risk for trades of size \(\Delta^k\), uniformly across all tiers, while the second term is tier-specific. This second term is nonlinear in inventory and highly sensitive to the presence and nature of skew sniffers in the tier: market makers need to only slightly reduce the skew shown to skew sniffers with high trading activity (as they contribute to global inventory risk reduction), whereas they must reduce it more substantially for those who trade infrequently.

As an illustration, consider a standard size ladder of \((1, 2, 5, 10, 20, 50)\) million (hereafter \(\text{M}\)) notional and two tiers with total baseline intensities \(\Lambda_0^{1, \cdot} + \Lambda_0^{2, \cdot} = (2000, 800, 600, 400, 120, 80)\)~day\(^{-1}\) and \(\kappa^{1,k} = \kappa^{2,k}  = 3\)~bp\(^{-1}\) for all~\(k \in  \mathcal K\). This corresponds to a relatively liquid currency pair with a daily turnover exceeding \(6\) billion notional and a top-of-book spread of approximately \(0.7\)~bp. We assume a daily volatility of \(100\)~bp and a market maker's risk aversion parameter of \(\gamma = 10^{-4}\)~bp\(^{-1}\)~M\(^{-1}\). For simulations, we use \(10^5\) Monte Carlo paths, a time horizon of \(T=10^4\) seconds, and the size ladder and intensity parameters stated above. In the reported simulations, the first-order formulas are evaluated at \(\varepsilon=1\) and quote offsets are floored at zero to prevent the local first-order formula from producing economically meaningless negative offsets when the approximation would otherwise be extrapolated too~far.

In the first tier, we assume no price reading (\(w^{1, k} = 0\), for all \(k \in \mathcal{K}\)), whereas in the second tier we assume the presence of skew sniffers with \(w^{2, k} = \frac{\exp(-\sqrt{\Delta^k/\Delta^1})}{1000}\), for all \(k \in \mathcal{K}\). We define the safe volume share as the ratio  \(\text{SVS} = \frac{\Lambda_0^{1, k}}{\Lambda_0^{1, k} + \Lambda_0^{2, k}}\), chosen identical across size indices \(k \in \mathcal{K}\). SVS is the proportion of trading intensity not subject to price reading.

\begin{figure}[H]\centering
\includegraphics[width=0.6\linewidth]{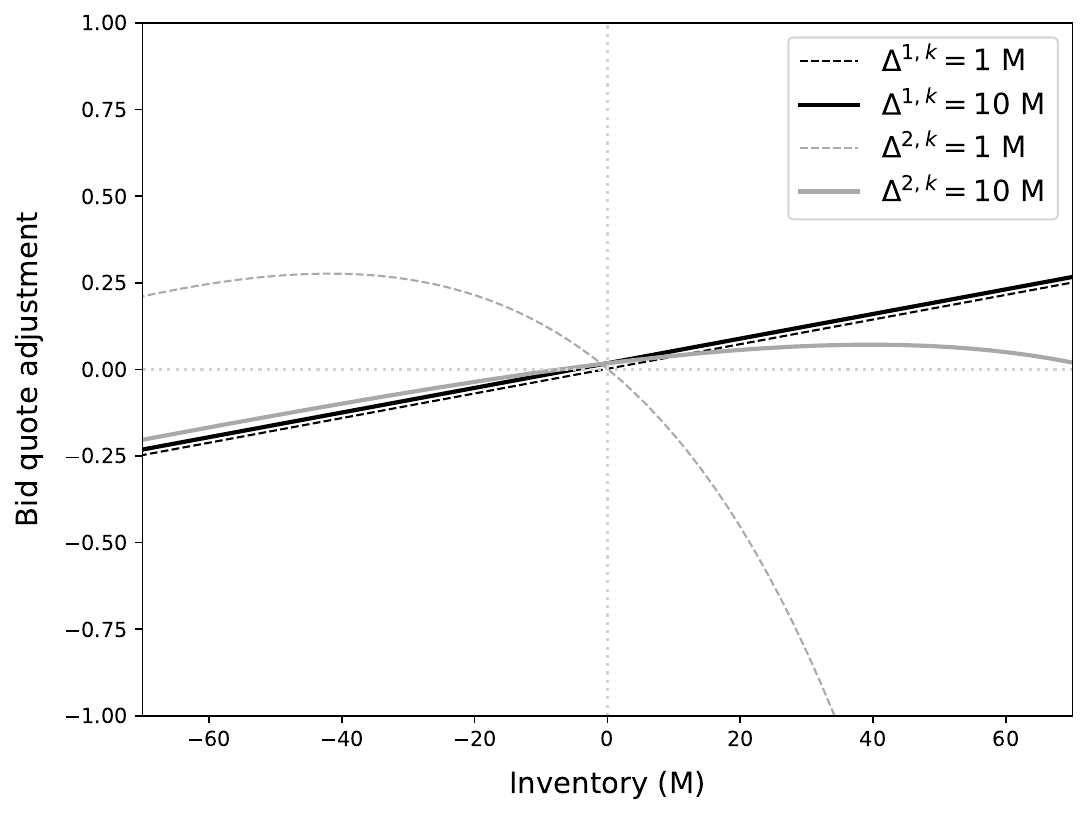}\\
\caption{Optimal bid quote adjustment,
\(\varepsilon^{-1} \left(\widehat{\mathfrak{d}}^{n,k,b*}(q) - \widehat{\delta}^{n,k,b*}(q) \right)\), with the second tier subject to price reading (we illustrate the first and the fourth sizes) -- \(\text{SVS} = 25\%\).}
\label{price_reading_a}
\end{figure}

\begin{figure}[H]\centering
\includegraphics[width=0.6\linewidth]{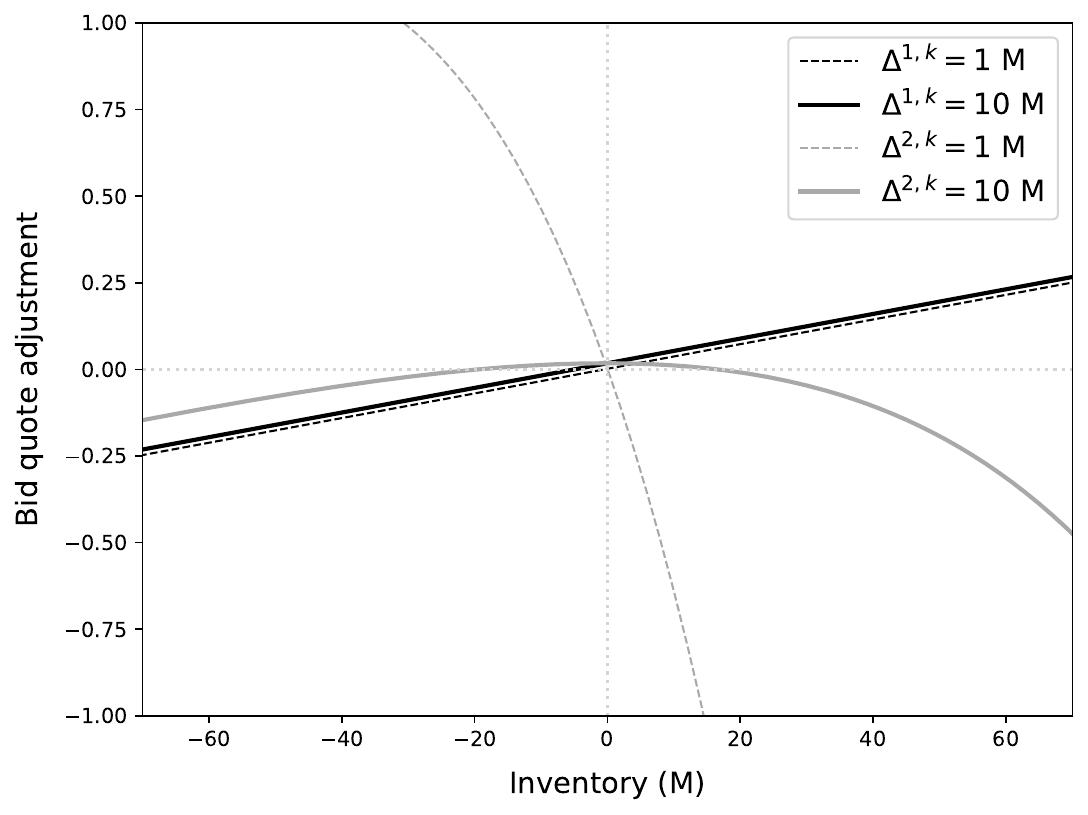}\\
\caption{Optimal bid quote adjustment,
\(\varepsilon^{-1} \left(\widehat{\mathfrak{d}}^{n,k,b*}(q) - \widehat{\delta}^{n,k,b*}(q) \right)\), with the second tier subject to price reading (we illustrate the first and the fourth sizes) -- \(\text{SVS} = 75\%\).}
\label{price_reading_b}
\end{figure}

As shown in Figures \ref{price_reading_a} and \ref{price_reading_b},\footnote{We only plot bid quote adjustments because ask quote adjustments are symmetric.} optimal pricing for the first tier reflects the global increase in informational risk: spread widening (more visible for large sizes) and an increase in skew. For the second tier, the additional skew related to the global increase in informational risk is counterbalanced by the objective to reduce information leakage, which implies a reduction in the skew shown to skew sniffers. In both figures, we clearly see that the skews are indeed lower for the second tier than for the first. Moreover, the de-skewing effect is particularly pronounced for top-of-book quotes in both the low and high \(\text{SVS}\) cases: the skew associated with top-of-book quotes for the second tier is not only lower than for the first tier, but it is in fact lower than in the absence of skew sniffers. The case of larger sizes is notable as it depends strongly on the value of \(\text{SVS}\). When \(\text{SVS} = 25\%\) (see Figure \ref{price_reading_a}), the skew shown to clients of the second tier for large sizes is still larger than in the absence of skew sniffers because they contribute significantly to risk reduction through their relatively high trading intensity. In contrast, when \(\text{SVS} = 75\%\) (see Figure \ref{price_reading_b}), the de-skewing effect dominates even for large sizes. In all cases, the contribution of the defensive side (here for positive inventories) to de-skewing is always higher than that of the offensive side (here for negative inventories).

Let us now simulate the PnL of a market maker facing price reading as described above, over the time window \([0,T]\), for different values of the safe volume share and several quoting strategies: (i) the ``No Action'' strategy, which does not take price reading into account and uses the quotes from Eq.~\eqref{num_quotes_no}, (ii) the ``No Skew'' strategy, which consists of withdrawing skew from prices shown to the tier of price readers, i.e., broadcasting zero-inventory pricing to clients of the second tier, and (iii) the ``Optimal'' strategy, which uses the first-order approximations of Eqs.~\eqref{num_quotes_1} and \eqref{num_quotes_2} with~\(\varepsilon=1\).

\begin{figure}[H]\centering
\includegraphics[width=0.6\linewidth]{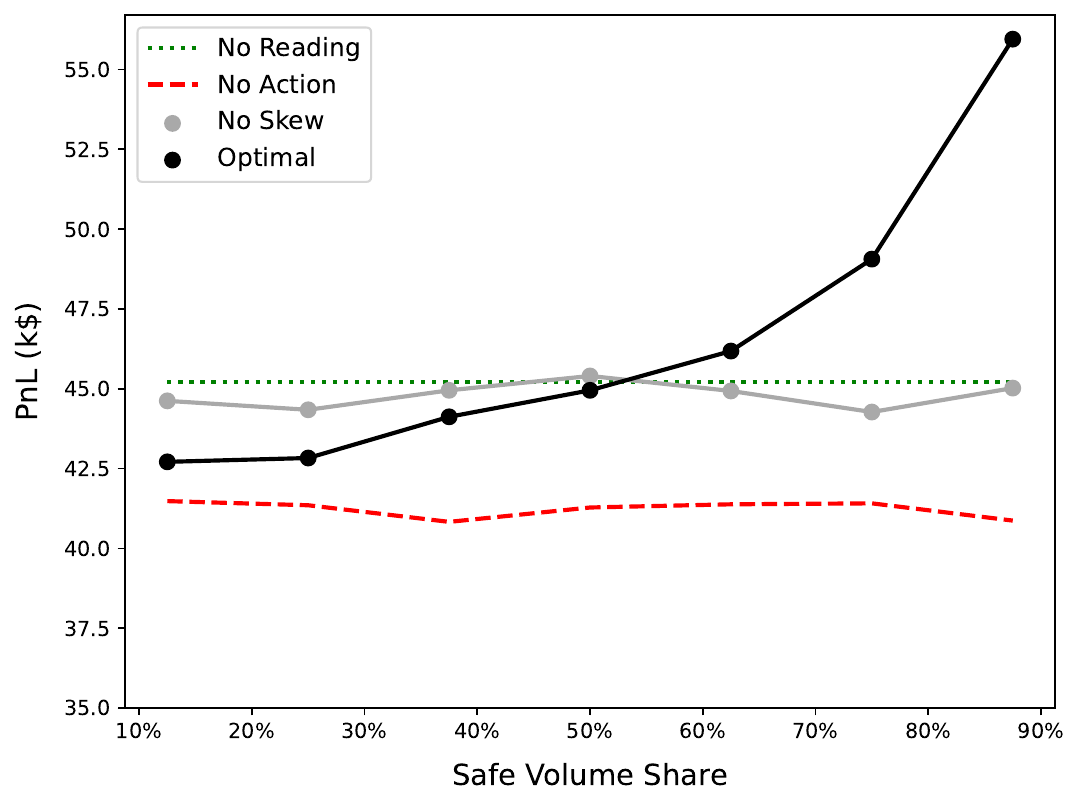}\\
\caption{Average values of the simulated market maker's PnL as a function of safe volume share for the different strategies.}
\label{price_reading_sim_pnl}
\end{figure}

\begin{figure}[H]\centering
\includegraphics[width=0.6\linewidth]{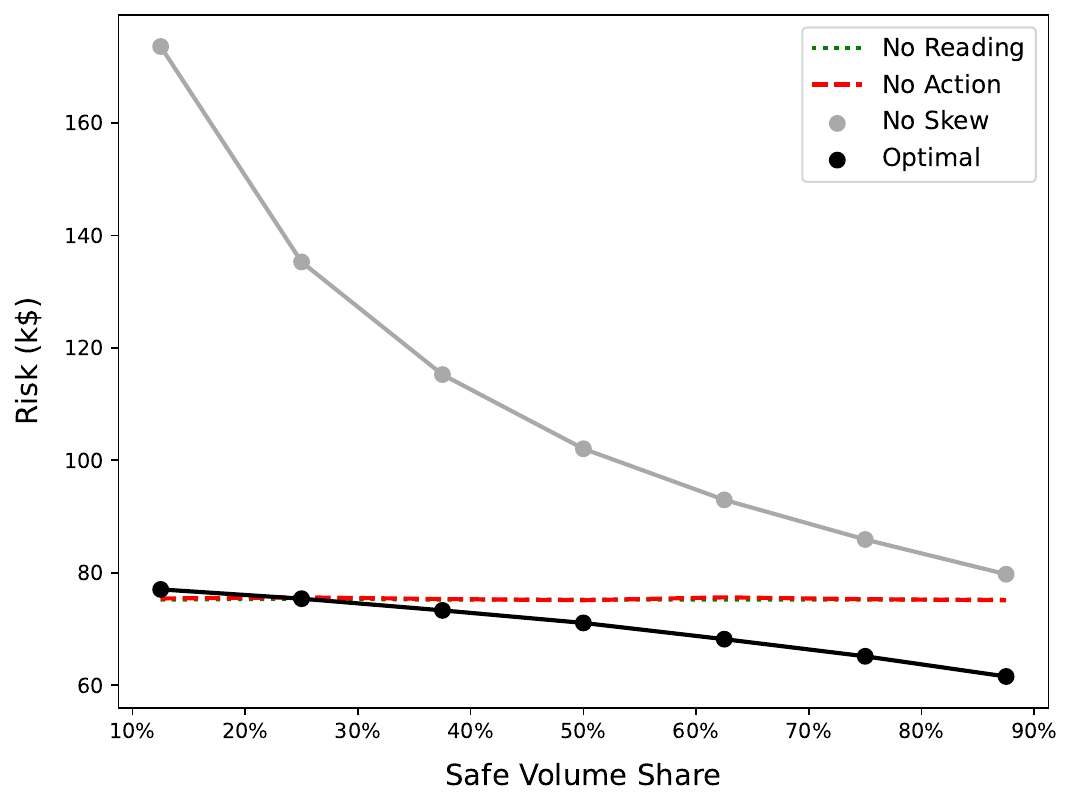}\\
\caption{Values of the standard deviation for the simulated market maker's PnL as a function of safe volume share for the different strategies.}
\label{price_reading_sim_risk}
\end{figure}

\begin{figure}[H]\centering
\includegraphics[width=0.6\linewidth]{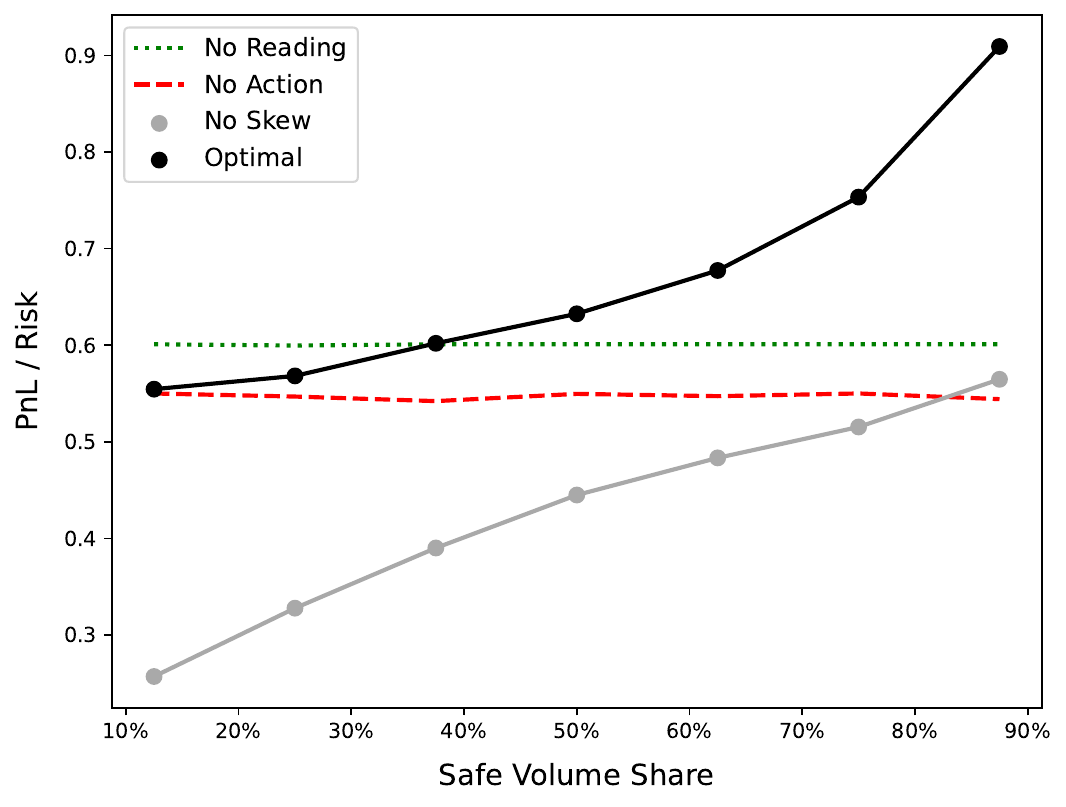}\\
\caption{Risk-adjusted performance measure associated with the simulated market maker's PnL as a function of safe volume share for the different strategies.}
\label{price_reading_sim_sharpe}
\end{figure}

Figures \ref{price_reading_sim_pnl}, \ref{price_reading_sim_risk} and \ref{price_reading_sim_sharpe} illustrate the average PnL, its standard deviation, and the ratio between the two (as a risk-adjusted performance measure), respectively, over the \(10^5\) independent simulations for several values of \(\text{SVS}\) when using the above three strategies. The ``No Reading'' counterfactual baseline corresponds to the absence of skew sniffers (\(w^{2,k} = 0\), for all \(k \in \mathcal K\)) and uses the quoting strategy of Eq.~\eqref{num_quotes_no}.

\vspace{5mm}

We clearly see that the presence of skew sniffers implies a decrease in the expected PnL compared to the baseline if quotes are not modified to account for price reading (``No Action'' strategy). However, the risk is not affected, and no dependence on the safe volume share is visible. The na{\"\i}ve ``No Skew'' approach (consisting of showing no skew to the skew sniffers) helps restore the PnL but leads to a considerable increase in risk, particularly if the price readers trade a lot (i.e., for low values of \(\text{SVS}\)). The pricing strategy based on our first-order approximations not only increases the PnL compared to that of the ``No Action'' strategy but also reduces the risk. This strategy is more efficient when the fraction of skew sniffers is low. For a very low safe volume share, de-skewing can be competitive in PnL/risk terms, but it reduces the dealer's access to risk-reducing flow and should be understood as a defensive benchmark rather than as a uniformly optimal response.\footnote{Note that a PnL higher than the baseline is not unexpected, as the strategy effectively introduces a controlled price drift.} The risk-adjusted performance measures of Figure \ref{price_reading_sim_sharpe} clearly emphasise the benefits of our first-order approximations.

\subsection{Adverse Selection}

Let us now consider the case of adverse selection alone. We have
\[
D^k_{\pm}\widehat{f}(q) = \frac{\sum_{\substack{m \in \mathcal{N} \\ j \in \mathcal{K}}} \alpha^{m,j} \Lambda_0^{m,j} (\beta^{m,j}-\kappa^{m,j})e^{\frac{\beta^{m,j}}{\kappa^{m,j}}}}{\sum_{\substack{m \in \mathcal{N} \\ j \in \mathcal{K}}} \Delta^j  \Lambda_0^{m,j} \kappa^{m,j}}\left(\pm q + \frac {\Delta^k}2 \right),
\]
and therefore the quote adjustment is given by
\begin{align*}
\frac{\widehat{\mathfrak{d}}^{n,k,b/a*}(q) - \widehat{\delta}^{n,k,b/a*}(q)}{\varepsilon}
&= \frac{\sum_{\substack{m \in \mathcal{N} \\ j \in \mathcal{K}}} \alpha^{m,j} \Lambda_0^{m,j} (\beta^{m,j}-\kappa^{m,j})e^{\frac{\beta^{m,j}}{\kappa^{m,j}}}}{\sum_{\substack{m \in \mathcal{N} \\ j \in \mathcal{K}}} \Delta^j  \Lambda_0^{m,j} \kappa^{m,j}}\left(\pm q + \frac {\Delta^k}2 \right)\\
&	\mp \frac{q \pm \Delta^k}{\Delta^k} \frac{\beta^{n,k} - \kappa^{n,k}}{\kappa^{n,k}} \zeta^{n,k}\left( \widehat\delta^{n,k,b/a*}(q) \right).
\end{align*}

As discussed above, the first term represents the global response to adverse selection risk, for trades of size \(\Delta^k\), uniformly across all tiers, whereas the second term is tier-specific. The magnitude and sign of the different effects depend critically on the relative slopes of the exponential functions defining the intensity and price impact functions -- that is, on the differences \(\beta^{m,j} - \kappa^{m,j}\) in the first sum, and on the tier-specific difference \(\beta^{n,k} - \kappa^{n,k}\) in the second term.

As an illustration, we consider the same market as in Section \ref{sec:price_reading} with a two-tier setup. The first tier is assumed to be standard, with \(\alpha^{1,k} = 0\) for all \(k \in \mathcal K\), while the second tier is endowed with a size-uniform informational advantage, characterised by \(\alpha^{2,k} = 0.05\) for all \(k \in \mathcal K\), and slow signals, with \(\beta^{2,k} = 2 < \kappa^{2,k}\) for all \(k \in \mathcal K\).

\begin{figure}[h!]\centering
\includegraphics[width=0.6\linewidth]{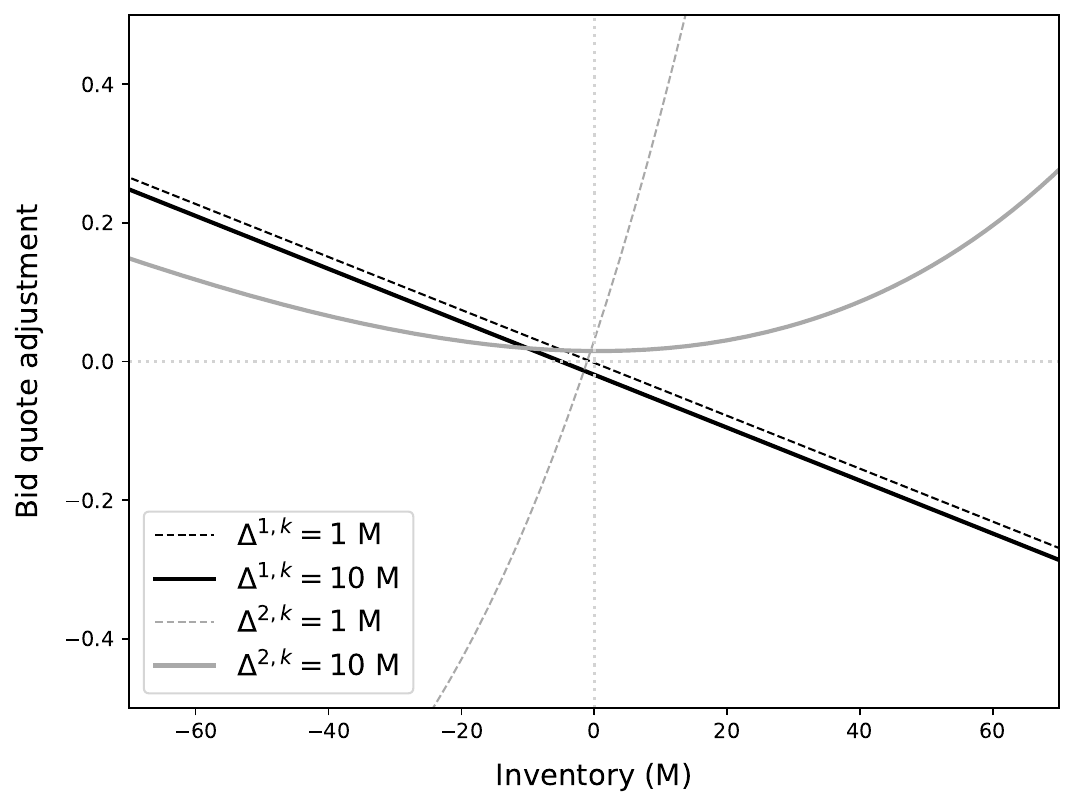}\\
\caption{Optimal bid quote adjustment,
\(\varepsilon^{-1} \left(\widehat{\mathfrak{d}}^{n,k,b*}(q) - \widehat{\delta}^{n,k,b*}(q) \right)\), with the second tier subject to adverse selection (we illustrate the first and the fourth sizes) -- \(\text{SVS} = 50\%\).}
\label{adverse_selection}
\end{figure}

Figure~\ref{adverse_selection} illustrates the impact of adverse selection on quotes when \(\text{SVS} = 50\%\). We see that, in the presence of adverse selection, the market maker behaves as if they were less risk averse when it comes to the quotes shown to the first tier. In particular, they tighten the bid-ask spread for the first tier when inventory is zero. By contrast, the spread quoted to the second (informed) tier is widened. Nevertheless, the market maker can still exploit the top-of-book prices shown to the informed tier to probe price direction. The limited effect on larger-size quotes is partly due to our modelling assumptions, but the underlying mechanism is general: small risk-reducing trades with informed clients cause limited losses while providing meaningful information. A key takeaway from this analysis is that a tier subject to adverse selection should not necessarily be avoided by the market maker contrary to what initial intuition might suggest. Rather, it can be leveraged as a form of ``signal subscription'' beneficial for the risk management of the entire franchise.

The picture is symmetrically reversed for adverse selection with sharp footprints (\(\beta^{2,k} > \kappa^{2,k}\), for all \(k \in \mathcal K\)) -- not shown. In this case, the market maker widens the spread for the first tier while top-of-book prices for risk-decreasing trades with the informed tier become less aggressive: the market maker can no longer extract valuable information from such trades and must instead focus on protection.

Let us now simulate the PnL of a market maker facing adverse selection as described above, over the same time window \([0,T]\) as above, for different values of the safe volume share and two quoting strategies: (i) the ``No Action'' strategy, which does not modify quotes to take adverse selection into account and therefore uses the quotes from Eq.~\eqref{num_quotes_no}, and (ii) the ``Optimal'' strategy, which uses the first-order approximations of Eqs.~\eqref{num_quotes_1} and \eqref{num_quotes_2} with~\(\varepsilon=1\).

\begin{figure}[H]\centering
\includegraphics[width=0.6\linewidth]{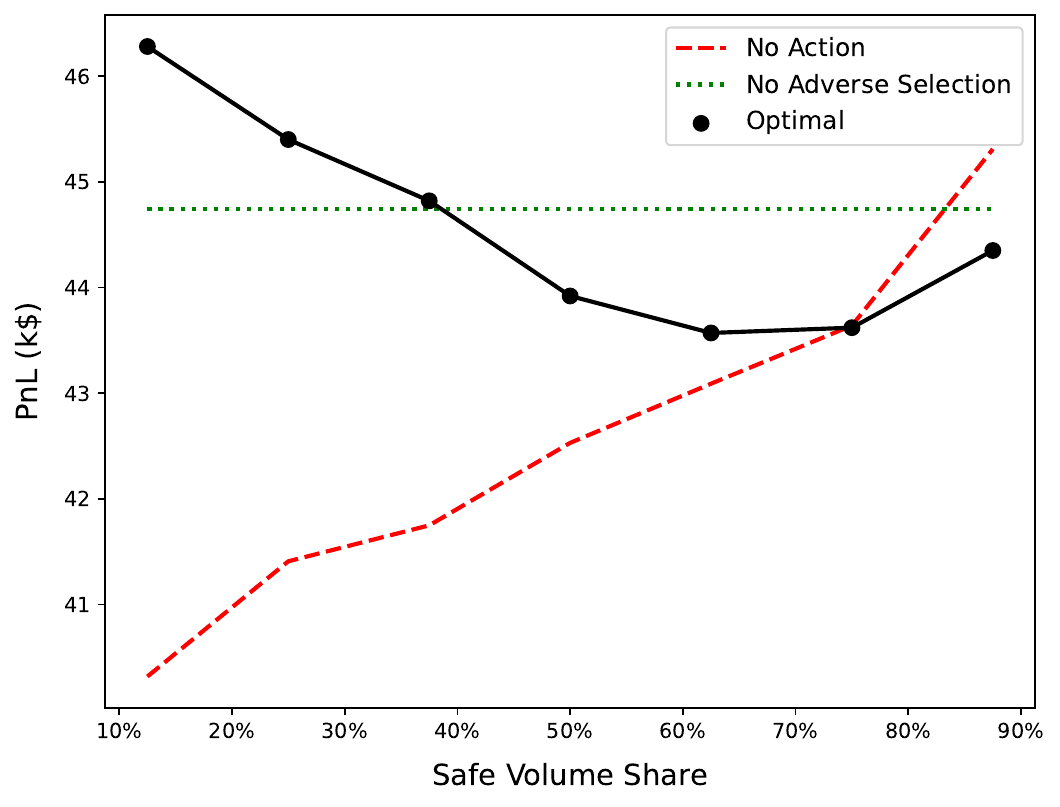}\\
\caption{Average values of the simulated market maker's PnL as a function of safe volume share, comparing the ``Optimal'' pricing strategy with the counterfactual baseline case
of no adverse selection and adverse selection with no strategic action by the market maker.}
\label{adverse_selection_sim_pnl}
\end{figure}

\begin{figure}[H]\centering
\includegraphics[width=0.6\linewidth]{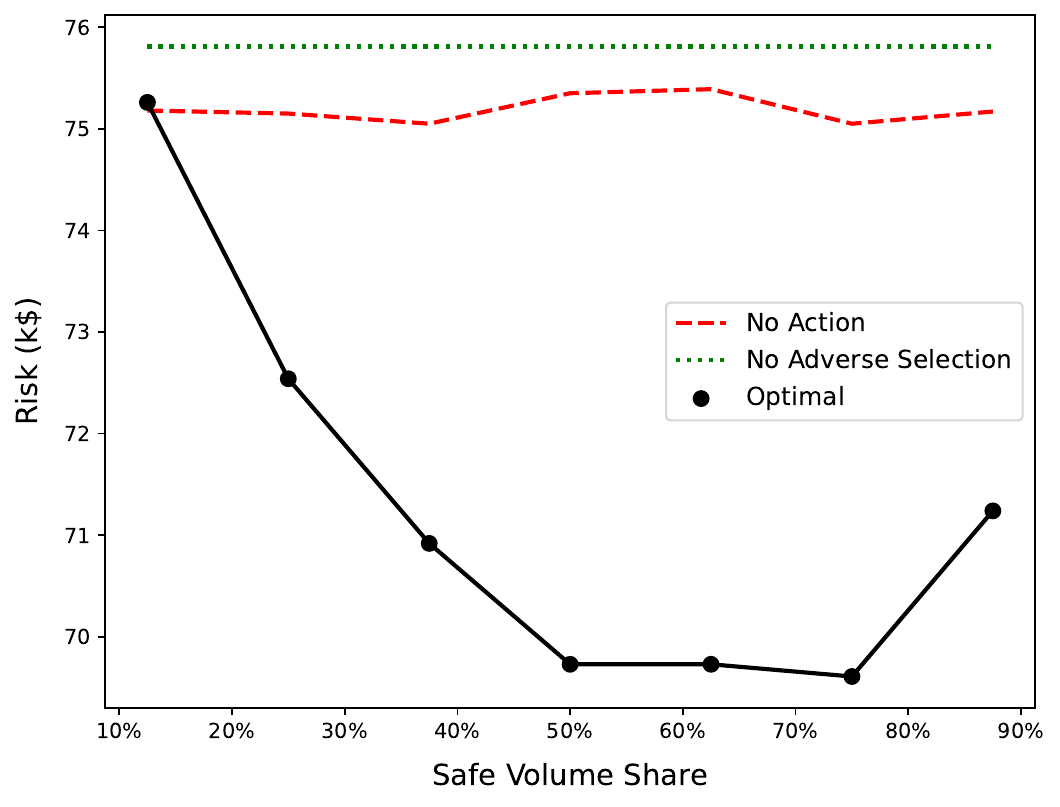}\\
\caption{Values of the standard deviation for the simulated market maker's PnL as a function of safe volume share, comparing the ``Optimal'' pricing strategy with the counterfactual baseline case of no adverse selection and adverse selection with no strategic action by the market maker.}
\label{adverse_selection_sim_risk}
\end{figure}

\begin{figure}[H]\centering
\includegraphics[width=0.6\linewidth]{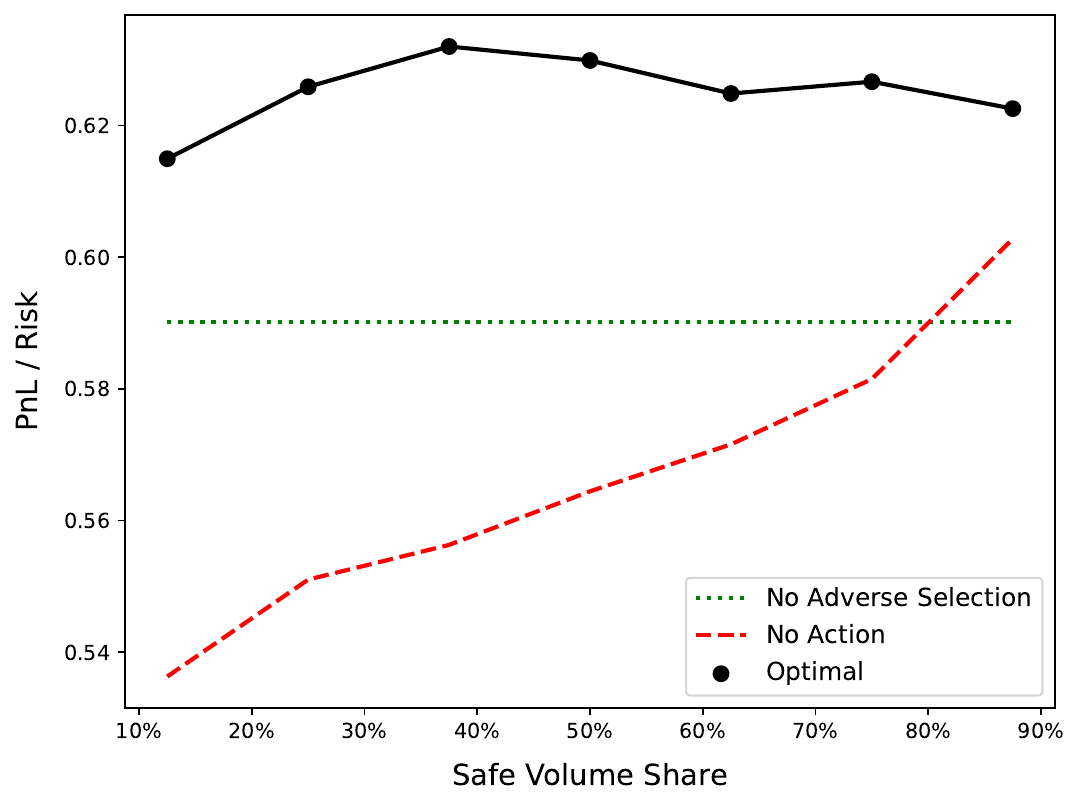}\\
\caption{Risk-adjusted performance measure associated with the simulated market maker's PnL as a function of safe volume share, comparing the ``Optimal'' pricing strategy with the counterfactual baseline case of no adverse selection and adverse selection with no strategic action by the market maker.}
\label{adverse_selection_sim_sharpe}
\end{figure}

Figures \ref{adverse_selection_sim_pnl}, \ref{adverse_selection_sim_risk}, and \ref{adverse_selection_sim_sharpe} illustrate the average PnL, its standard deviation, and the ratio between the two (as a risk-adjusted performance measure), respectively, over the \(10^5\) independent simulations for several values of \(\text{SVS}\) when using the above two strategies. The ``No Adverse Selection'' counterfactual baseline corresponds to the absence of adverse selection (\(\alpha^{2,k} = 0\), for all \(k \in \mathcal K\)) and uses the quoting strategy of Eq.~\eqref{num_quotes_no}.\footnote{Unlike in the case of price reading, there is no obvious na{\"\i}ve strategy to use for comparison in the case of adverse selection. The only alternative would be to stop pricing to the suspected tier, which would drastically damage the PnL and make any comparison irrelevant.} We clearly see that adverse selection negatively impacts the expected PnL while the risk remains almost identical (up to a slight decrease) when no action is taken. The use of our first-order approximations of the optimal pricing strategy helps to protect the expected PnL and significantly reduces risk, particularly when the share of informed clients is moderate. The risk-adjusted performance measures in Figure \ref{adverse_selection_sim_sharpe} clearly emphasise the benefits of our first-order approximations, with no clear dependence on~\(\text{SVS}\).

\section{Conclusion}
\label{sec:conclusion}

In this article, we have proposed a tractable stochastic optimal control
framework that enables market makers to adjust their quotes to better account
for informational risk arising from both price reading and adverse selection.
Using perturbation analysis, we have found that the impact of informational risk
can be split into two components: (i) a global component, related to induced
changes in the value function, that applies to pricing for all clients, and (ii)
a tier-specific component, aimed at mitigating the informational risk associated
with each tier.

In the presence of price reading, the global component leads the market maker to
act as if they were more risk averse when showing prices to their global
franchise. The tier-specific component pushes in the opposite direction for the
tiers concerned, as the market maker reduces the informational content of the
quotes shown to skew sniffers. Which of the two dominates depends on the
contribution of price readers to the market maker's risk reduction: skew
sniffers who trade actively still earn a significant skew, whereas those who
trade infrequently see it substantially withdrawn.

The optimal response to adverse selection depends on the relative shape of the
informed flow intensity and price impact functions. The market maker can become
less risk averse for their global franchise when informed clients trade on
long-term signals of the order of several dozen basis points to a few percentage
points. Top-of-book prices for risk-reducing trades then become very attractive
for these informed clients: this is the cost the market maker is willing to pay
to improve the risk management of their overall portfolio. In contrast, adverse selection with sharp signals of a few basis points forces the market maker to act protectively, showing defensive quotes even for risk-reducing trades with informed clients and behaving as if they were more risk averse with their global franchise.

\section*{Appendix}

The linear price-reading specification of Section~\ref{sec:genpb} is
analytically convenient and appropriate for a first-order expansion around
competitive quotes. Taken literally, however, it attributes informational
content to quotes that are too far from the market to be read at all. A bounded
alternative is to replace the raw skew by a smooth competitiveness score: for
each tier \(n\) and size \(k\), let
\[
s^{n,k,b/a}(\delta)
=
\frac{1}{1+\exp\left((\delta-\bar\delta^{n,k,b/a})/\tau^{n,k,b/a}\right)},
\]
where \(\bar\delta^{n,k,b/a}\) is a reference offset -- for instance the
baseline quote at zero inventory -- and \(\tau^{n,k,b/a}>0\) controls the
transition between competitive and noncompetitive quotes. The price-reading
signal of tier \(n\) is then
\[
I^n\left(\delta^{n,\cdot,b},\delta^{n,\cdot,a}\right)=
\sum_{k\in\mathcal K}w^{n,k}
\left(
s^{n,k,a}(\delta^{n,k,a})
-
s^{n,k,b}(\delta^{n,k,b})
\right),
\]
and the informational drift is modelled through \(\tilde{J}^n\left(I^n\left(\delta^{n,\cdot,b},\delta^{n,\cdot,a}\right)\right)\). This
signal is bounded by construction: a side of the ladder contributes to price
reading only insofar as it remains competitive enough to be informative.

\begin{figure}[H]\centering
\includegraphics[width=0.6\linewidth]{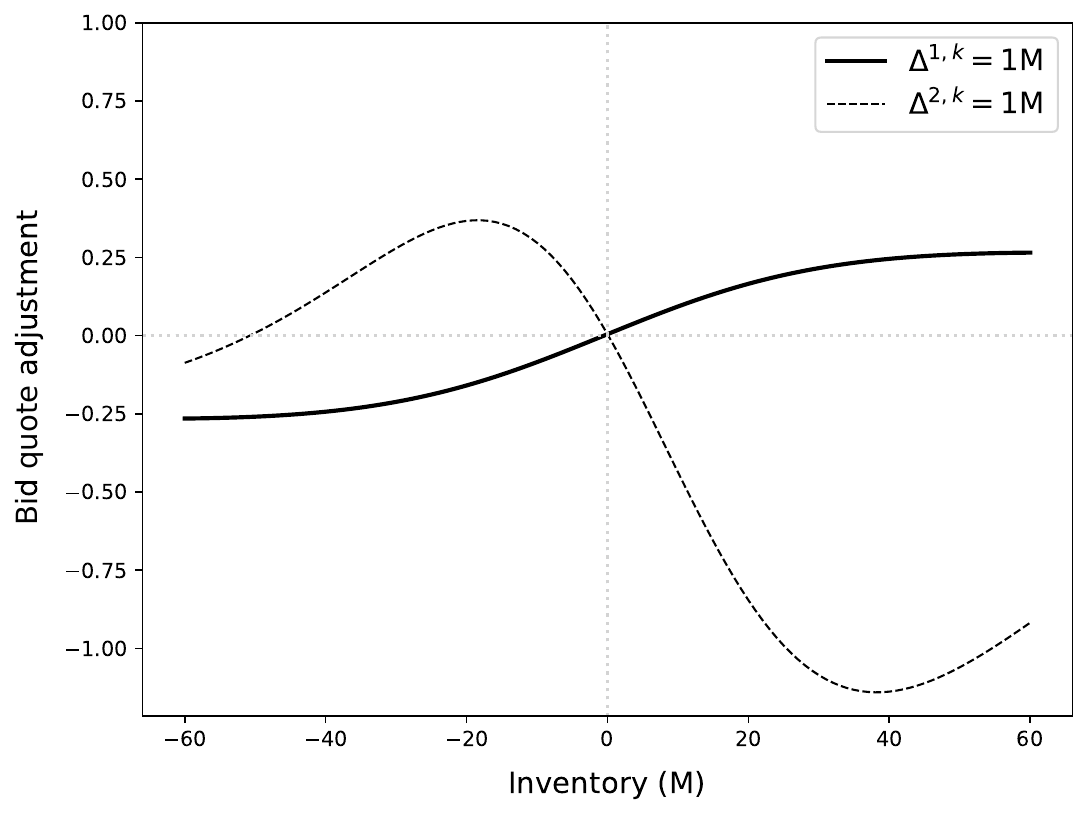}\\
\caption{Optimal normalised top-of-book bid quote adjustment for the logistic
competitiveness-based price-reading model, with \(\tau^{n,k,b/a}=0.1\). Other
parameters are as in Section~\ref{sec:price_reading}.}
\label{price_reading_competitive_appendix}
\end{figure}

The baseline problem is unchanged, so the first-order expansion of
Section~\ref{sec:foe} goes through verbatim; only the local price-reading
sensitivities are modified, \(w^{n,k}J^{n\,\prime}\) being replaced by the
derivatives of the competitiveness scores at the baseline quotes. The separation
between a global value function correction and a local tier-specific quote
correction is therefore preserved. Figure~\ref{price_reading_competitive_appendix}
confirms that the mechanisms described in Section~\ref{sec:price_reading} are
recovered for moderate inventories, while extreme quotes no longer generate
unreasonable informational effects.

\section*{Acknowledgements}

The authors would like to express their sincere gratitude to all those who made this research possible, and especially to Richard Anthony (HSBC) and Jean-Michel Beacco (Institut Louis Bachelier). The authors' thoughts go to Jean-Michel, who sadly died in a mountaineering accident while this paper was being written.

\section*{Statement on Funding}

The research conducted for this paper benefited from the financial support of the HSBC FX Research Initiative, a Research Initiative under the aegis of Institut Louis Bachelier, in partnership with HSBC UK. The views expressed are those of the authors and do not necessarily reflect the views or practices at HSBC.

\section*{Declaration of Generative AI Use}

During the preparation of this manuscript, the authors used various versions of OpenAI's ChatGPT, Google's Gemini and Anthropic's Claude to assist with editing, proofreading and syntactic refinement, with a view to improving the readability and language quality of the paper. These tools did not contribute to the original ideas, the analysis or the conclusions of this work. The authors reviewed and edited all content after using them and take full responsibility for the final text -- including any misspellings they may have reintroduced along the way.

\section*{Statement on Data Availability}

The data supporting the findings of this study were generated via numerical computations and simulations. The models, methods and parameters used to produce these data are fully described within the manuscript, enabling the reproduction of the results. The code used for this study is not publicly deposited. Reasonable inquiries regarding the replication of the findings may be directed to the authors via email.


\begin{thebibliography}{10}

\bibitem{avellaneda2008high}
Marco Avellaneda and Sasha Stoikov.
\newblock High-frequency trading in a limit order book.
\newblock {\em Quantitative Finance}, 8(3):217--224, 2008.

\bibitem{baldauf2024}
Markus Baldauf and Joshua Mollner. 
\newblock Competition and Information Leakage. 
\newblock {\em Journal of Political Economy}, 132(5):1603--1641, 2024.

\bibitem{barzykin2023algorithmic}
Alexander Barzykin, Philippe Bergault, and Olivier Guéant.
\newblock Algorithmic market making in dealer markets with hedging and market impact.
\newblock {\em Mathematical Finance}, 33(1):41--79, 2023.

\bibitem{barzykin2023dealing}
Alexander Barzykin, Philippe Bergault, and Olivier Guéant.
\newblock Dealing with multi-currency inventory risk in foreign exchange cash markets.
\newblock {\em Risk}, March 2023.

\bibitem{barzykin2024market}
Alexander Barzykin, Philippe Bergault, and Olivier Guéant.
\newblock Market-making in spot precious metals.
\newblock {\em Risk}, December 2024.

\bibitem{bergault2021closed}
Philippe Bergault, David Evangelista, Olivier Guéant, and Douglas Vieira.
\newblock Closed-form approximations in multi-asset market making.
\newblock {\em Applied Mathematical Finance}, 28(2):101--142, 2021.

\bibitem{butz2019}
Maximilian Butz and Roel Oomen.
\newblock Internalisation by electronic FX spot dealers.
\newblock {\em Quantitative Finance}, 19(1):35--56, 2019.

\bibitem{cartea2015algorithmic}
{\'A}lvaro Cartea, Sebastian Jaimungal, and Jos{\'e} Penalva.
\newblock {\em Algorithmic and High-Frequency Trading}.
\newblock Cambridge University Press, 2015.

\bibitem{cartea2014buy}
{\'A}lvaro Cartea, Sebastian Jaimungal, and Jason Ricci.
\newblock Buy low, sell high: A high frequency trading perspective.
\newblock {\em SIAM Journal on Financial Mathematics}, 5(1):415--444, 2014.

\bibitem{cartea2025simple}
{\'A}lvaro Cartea and Leandro Sánchez-Betancourt.
\newblock A Simple Strategy to Deal with Toxic Flow.
\newblock {\em arXiv preprint arXiv:2503.18005}, 2025.

\bibitem{glosten1985bid}
Lawrence R. Glosten and Paul R. Milgrom.
\newblock Bid, ask and transaction prices in a specialist market with heterogeneously informed traders.
\newblock {\em Journal of Financial Economics}, 14(1):71--100, 1985.

\bibitem{goyder2024}
Bernard Goyder and Cole Lipsky.
\newblock BNPP ups efforts to weed out skew sniffers.
\newblock {\em FX Markets}, 20 Nov 2024.

\bibitem{gueant2016financial}
Olivier Gu{\'e}ant.
\newblock {\em The Financial Mathematics of Market Liquidity: From Optimal Execution to Market Making}, volume 33.
\newblock CRC Press, 2016.

\bibitem{gueant2013dealing}
Olivier Gu{\'e}ant, Charles-Albert Lehalle, and Joaquin Fernandez-Tapia.
\newblock Dealing with the inventory risk: a solution to the market making problem.
\newblock {\em Mathematics and Financial Economics}, 7(4):477--507, 2013.

\bibitem{ho1981optimal}
Thomas Ho and Hans~R. Stoll.
\newblock Optimal dealer pricing under transactions and return uncertainty.
\newblock {\em Journal of Financial Economics}, 9(1):47--73, 1981.

\bibitem{lucchese2024}
Lorenzo Lucchese, Mikko S.~Pakkanen, and Almut E.D.~Veraart.
\newblock The short-term predictability of returns in order book markets: A deep learning perspective.
\newblock {\em International Journal of Forecasting}, 40(4):1587--1621, 2024.

\bibitem{rosenbaum2022}
Mathieu Rosenbaum and Jianfei Zhang.
\newblock Multi-asset market making under the quadratic rough Heston.
\newblock {\em arXiv:2212.10164v1}, 2022.

\bibitem{sirignano2019}
Justin Sirignano and Rama Cont. 
\newblock Universal features of price formation in financial markets: perspectives from deep learning. 
\newblock {\em Quantitative Finance}, 19(9):1449--1459, 2019.


\end{thebibliography}
\end{document}